\documentclass[a4paper,fleqn,usenatbib]{mnras}

\usepackage{mathptmx}

\usepackage[T1]{fontenc}
\usepackage{ae,aecompl}

\usepackage{graphicx}	
\usepackage{amsmath}	
\usepackage{amssymb}	

\title[The {\it{Gaia}}-ESO Survey: Milky Way field targets selection]{The {\it{Gaia}}-ESO Survey: the selection function of the Milky Way  field stars}

\author[Edita Stonkut{\.e} et al.]{
E.~Stonkut{\.e},$^{1,2}$\thanks{{\tt email: edita@astro.lu.se}}
S.~E.~Koposov,$^{3}$ 
L.~M.~Howes,$^{1}$
S.~Feltzing,$^{1}$
C.~C.~Worley,$^{3}$
\newauthor
G.~Gilmore,$^{3}$
G.~R.~Ruchti,$^{1}$
G.~Kordopatis,$^{4}$
S. Randich,$^{5}$
T. Zwitter,$^{6}$
T.~Bensby,$^{1}$
\newauthor
A. Bragaglia,$^{7}$
R. Smiljanic,$^{8}$
M.~T. Costado,$^{9}$
G. Tautvai\v{s}ien\.{e}$^{2}$
A.~R. Casey,$^{3}$
\newauthor
A.~J. Korn,$^{10}$
A.~C. Lanzafame,$^{11,12}$
E. Pancino,$^{5,13}$
E. Franciosini,$^{5}$
A. Hourihane,$^{3}$
\newauthor
P. Jofr\'e,$^{3}$
C. Lardo,$^{14}$
J. Lewis,$^{3}$
L. Magrini,$^{5}$
L. Monaco,$^{15}$
L. Morbidelli,$^{5}$
\newauthor
G.~G. Sacco,$^{5}$
L. Sbordone,$^{16,17}$
\\
$^{1}$Lund Observatory, Department of Astronomy and Theoretical Physics, Box 43, SE-22100, Lund, Sweden\\
$^{2}$Institute of Theoretical Physics and Astronomy, Vilnius University, Saul\.{e}tekio al. 3, LT-10222, Vilnius, Lithuania\\
$^{3}$Institute of Astronomy, Cambridge University, Madingley Road, Cambridge CB3 0HA, United Kingdom\\
$^{4}$Leibniz-Institut f\"ur  Astrophysik Potsdam (AIP), An der Sternwarte 16, 14482 Potsdam, Germany\\
$^{5}$INAF - Osservatorio Astrofisico di Arcetri, Largo E. Fermi 5, 50125, Florence, Italy\\
$^{6}$Faculty of Mathematics and Physics, University of Ljubljana, Jadranska 19, 1000, Ljubljana, Slovenia\\
$^{7}$INAF - Osservatorio Astronomico di Bologna, via Ranzani 1, 40127, Bologna, Italy\\
$^{8}$Nicolaus Copernicus Astronomical Center, Polish Academy of Sciences, ul. Bartycka 18, 00-716, Warsaw, Poland\\
$^{9}$Instituto de Astrof\'{i}sica de Andaluc\'{i}a-CSIC, Apdo. 3004, 18080 Granada, Spain\\
$^{10}$Department of Physics and Astronomy, Uppsala University, Box 516, SE-751 20 Uppsala, Sweden\\
$^{11}$Universit\`a di Catania, Dipartimento di Fisica e Astronomia, Sezione Astrofisica, Via S. Sofia 78, I-95123 Catania, Italy \\
$^{12}$INAF-Osservatorio Astrofisico di Catania, Via S. Sofia 78, I-95123 Catania, Italy\\
$^{13}$ASI Science Data Center, via del Politecnico SNC, I-00133 Roma, Italy\\
$^{14}$Astrophysics Research Institute, Liverpool John Moores University, 146 Brownlow Hill, Liverpool L3 5RF, United Kingdom\\
$^{15}$Departamento de Ciencias Fisicas, Universidad Andres Bello, Republica 220, Santiago, Chile\\
$^{16}$Millennium Institute of Astrophysics, Av. Vicu\~{n}a Mackenna 4860, 782-0436 Macul, Santiago, Chile\\
$^{17}$Pontificia Universidad Cat\'{o}lica de Chile, Av. Vicu\~{n}a Mackenna 4860, 782-0436 Macul, Santiago, Chile\\
}

\date{Accepted 2016 April 25. Received 2016 April 25; in original form 2016 March 17}

\pubyear{2016}

\begin{document}
\label{firstpage}
\pagerange{\pageref{firstpage}--\pageref{lastpage}}
\maketitle

\begin{abstract}
The {\it{Gaia}}-ESO Survey was designed to target all major Galactic components (i.e., bulge, thin and thick discs, halo and clusters), with the goal of constraining the chemical and dynamical evolution of the Milky Way. 
This paper presents the methodology and considerations that drive the selection of the targeted, allocated and successfully observed Milky Way {\it{field stars}}. The detailed understanding of the survey construction, specifically the influence of target selection criteria on observed Milky Way field stars is required in order to analyse and interpret the survey data correctly. 
We present the target selection process for the Milky Way field stars observed with VLT/FLAMES and provide the weights that characterise the survey target selection. 
The weights can be used to account for the selection effects in the {\it{Gaia}}-ESO Survey data for scientific studies. We provide a couple of simple examples to highlight the necessity of including such information in studies of the stellar populations in the Milky Way.
\end{abstract}




\section{Introduction}\label{sec-intro}

The Milky Way is just one of hundreds of billions of galaxies that populate our visible Universe.

However, it is the one galaxy that we can study in the greatest detail. For example, thanks to spectroscopic surveys over the last few decades our understanding of the chemical evolution of our own Galaxy has increased
tremendously (for reviews see \citealt{freeman02, feltzing13, blandHawthorn16}).
A number of large-scale spectroscopic surveys of stars in the Milky Way have been completed or are underway, e.g., SEGUE \citep{yanny09}, RAVE \citep{steinmetz06}, GALAH \citep{deSilva15}, APOGEE \citep{majewski15}, LAMOST \citep{deng12}, {\it{Gaia}}-ESO \citep{gilmore12, randich13}, {\it{Gaia}} \citep{perryman01} or are being planned e.g. WEAVE \citep{dalton12}, MOONS \citep{cirasuolo12} and 4MOST \citep{deJong14}.
These are opening a new path to study formation and evolution of the Galaxy in great detail.  
All spectroscopic surveys of the Milky Way will suffer from selection effects. 
For example the object targeting algorithm employed in the survey will cause selection biases. Therefore we need to design our target selection algorithms to be as simple as possible.
This way we can determine how the observed spectroscopic sample represents the stars in the parent stellar population. All selection effects need to be accounted for when we want to
extrapolate from the observed volume to the ``global'' volume of the Milky Way.

There have been several SEGUE papers that have demonstrated the importance of accounting for the observational biases in different SEGUE samples. 
\citet{cheng12} examined the observational biases of the main sequence turn-off stars on low-latitude plates and they stress the importance of the weighting procedure for the proper correction
for selection biases. Furthermore, \citet{schlesinger12} determined and corrected for the effect of the SEGUE target selection on cool dwarf stars (G- and K type). A portion of this sample was also studied and corrected for biases in a different way by
\citet{bovy12} and \citet{liu12}. Selection effects are also considered in other analyses of spectroscopic survey data \citep[e.g. RAVE, APOGEE,][]{francis13, nidever14}. 
In this context it is important to discuss the {\it{Gaia}}-ESO Survey construction: how targets are selected; allocated on the spectrograph; and finally -- successfully observed. 

The {\it{Gaia}}-ESO Survey observing strategy has been constructed to answer specific scientific questions. The full survey
includes all major stellar populations: the Galactic inner and outer bulge, inner and outer thick and thin discs, the halo, currently known halo
streams, and star clusters. Selected targets consist of early- and late- type stars, metal-rich and metal-poor stars, dwarfs, giants, and cluster stars across the evolutionary sequence selected from previous studies of open clusters.

By the end, the survey will have observed with FLAMES/UVES a sample of several thousand FG-type stars within 2~kpc of the Sun in order to derive the detailed kinematic and elemental abundance distribution functions of the solar neighbourhood. 
The sample includes mainly thin and thick disc stars, of all ages and metallicities, but also a small fraction of local halo stars. FLAMES/GIRAFFE will observe a statistically significant  ($\sim$~10$^5$) sample of stars in all major stellar populations.

The {\it{Gaia}}-ESO Survey will provide a legacy dataset that adds great merit to the astrometric {\it{Gaia}} space mission by assembling a catalog of representative spectra for stars which {\it{Gaia}} will deliver  highly accurate proper motions but not detailed spectroscopic information. These combined data will allow us to probe for example the properties of the Galactic disc by looking for traces of past, and ongoing, accretion events.

While the {\it{Gaia}}-ESO Survey is currently still completing the observing campaign, there are scientific questions that are already being answered. These cover testing the nature of the thick disc and its relation to the thin disc 
\citep{recioblanco14, mikolaitis14, kordopatis15}; studying the relationship between age and metallicity, and the spatial distribution of stars \citep{bergemann14}; identifying the remnants of ancient building blocks of the Milky Way \citep{ruchti15};
determining the chemical composition of recently discovered ultra-faint satellites \citep{koposov15}; analysing metal-poor stars \citep{howes14, jacksonjones14};
and determining the chemical abundance distribution in globular and open clusters \citep{donati14, magrini15, sanroman15, tautvaisiene15}.

In this paper, we present the target selection process only for the Milky Way field stars observed in the {\it{Gaia}}-ESO Survey and provide the weights that characterise the survey sample. 

The {\it{Gaia}}-ESO is a public survey and the stellar spectra are available after observations, while reduced spectra and the astrophysical results obtained by the {\it{Gaia}}-ESO analysis teams are available to the general community via public releases through the ESO data archive\footnote{http://archive.eso.org/wdb/wdb/adp/phase3\_spectral/\\form?phase3\_collection=GaiaESO}.

This paper is organized as follows. In Section~\ref{sec:observ_setup} we introduce the observational setup of the {\it{Gaia}}-ESO Survey. Section~\ref{sec:methods} describes the methods used to select targets for the Milky Way field observations with FLAMES/GIRAFFE and FLAMES/UVES. 
The initial target selection for GIRAFFE is presented in Section~\ref{sec:Giraffe} and for UVES in Section~\ref{sec:UVES}. In Section~\ref{sec:Final} we introduce the final target selection and in Section~\ref{sec:Weights} the weights used to correct for selection effects, calculated after target selection and allocation. In Section~\ref{sec:idr4} we take a first look at the {\it{Gaia}}-ESO Survey iDR4 data and discuss the completeness of the successfully analysed spectroscopic sample. We provide a simple example  of the metallicity distribution and how it is affected by the selection effects. We show that the metallicity distribution of the Milky Way field stellar sample observed in the {\it{Gaia}}-ESO Survey can be corrected to a distribution unaffected by the selection bias by applying the calculated weights.
 Finally, in Section~\ref{sec:Conclusions} we discuss the implications of our results and give concluding remarks.


\section{Observational setup for Milky Way fields}\label{sec:observ_setup}

\begin{figure}
\includegraphics[width=0.8\columnwidth]{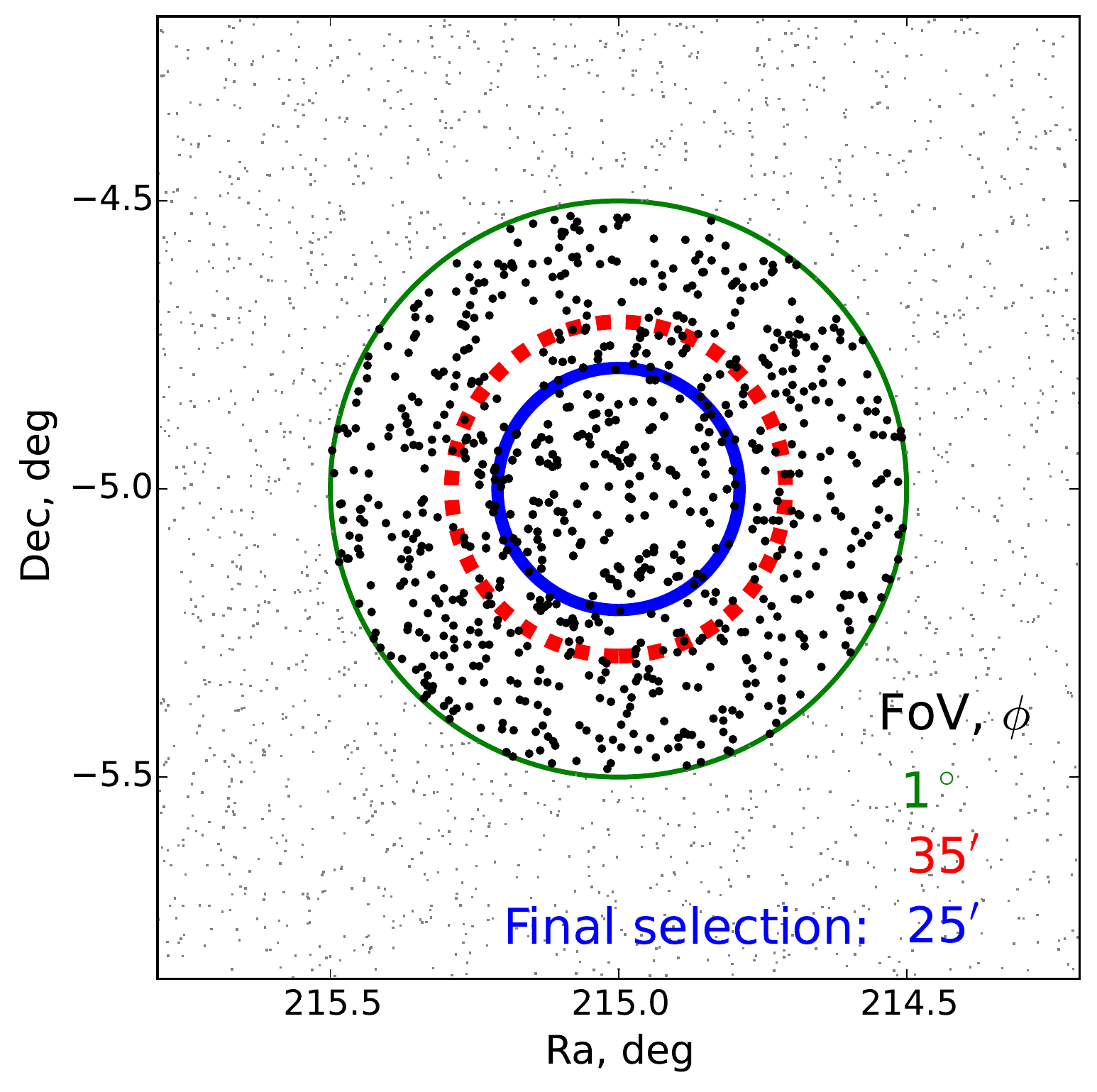}
\caption{The plot shows stars in one of the {\it{Gaia}}-ESO Milky Way field, GES\_MW\_142000-050000. Small dots are surrounding stars while the large dots are stars involved in the calculation of the selection function (see Section \ref{sec:Final} for details). Green thin line -- a 1-degree (diameter) field-of-view; red dashed line -- a 35 arcmin (diameter) field-of-view;
blue solid line -- final target selection with a 25 arcmin (diameter) field-of-view (i.e. that at FLAMES on VLT).}
\label{fig-FoV}
\end{figure}

The observations are conducted with the Fibre Large Array Multi Element Spectrograph (FLAMES) \citep{pasquini02} at the Very Large Telescope array (VLT) operated by the European Southern Observatory on Cerro Paranal, Chile.
FLAMES is a fibre facility of the VLT and is mounted at the Nasmyth A platform of the second Unit Telescope of VLT. This instrument has a large 25 arcmin diameter field-of-view (see Fig.~\ref{fig-FoV}).

One of the three main FLAMES components is a Fibre Positioner (OzPoz) which hosts two plates. While one plate is observing the other plate is configuring fibres so that they are positioned for the subsequent observation. This limits the overhead between one observation and the next to less than 15 minutes, including the telescope preset and the acquisition of the next field.
The fibre facility is equipped with two sets of 132 and 8 fibres to feed two different spectrographs GIRAFFE and UVES, respectively.

The medium resolution spectrograph GIRAFFE with the two setups -- HR10 ($\lambda$=533.9-561.9 nm, R$\sim$19800) and HR21 ($\lambda$=848.4-900.1 nm, R~$\sim$~16200) was used to observe the Milky Way field stars. 
To observe Milky Way field stars in high-resolution mode the survey used the UV-Visual Echelle Spectrograph (UVES) with a setup centered at 580 nm ($\lambda$=480-680 nm, R$\sim $47000) 
\citep{dekker00}. 

\begin{figure*}
\includegraphics[width=1.3\columnwidth]{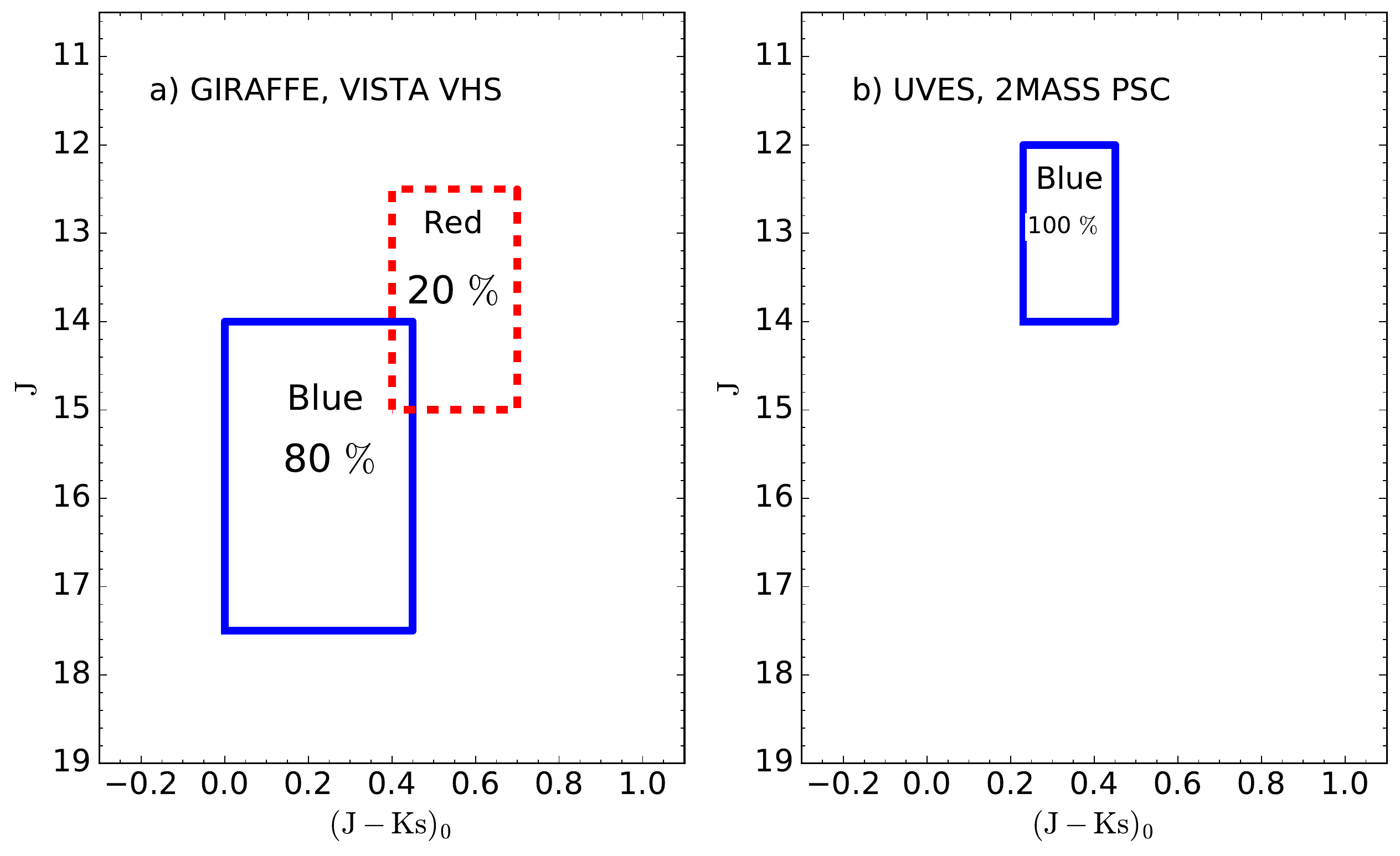}
\caption{The basic colour-magnitude schemes for target selection (a) GIRAFFE  and (b) UVES,  respectively. Blue solid line -- shows the area from which targets are assigned to the blue box, and the red dashed line -- to the red box. The selection of targets is 
based on VISTA VHS photometry and 2MASS photometry as indicated. On the x-axis we show the de-reddened (J-K$_{s}$)$_0$ colour, whereas on the y-axis we show the observed $J$ magnitude (for more details see Section~\ref{subsec:Reddening}).}
\label{fig-main_boxes}
\vspace{0mm}
\end{figure*}

\section{TARGET SELECTION METHODS}\label{sec:methods} 

The {\it{Gaia}}-ESO Survey is designed to select and observe three classes of targets in the Milky Way -- field stars, candidate members of open clusters and calibration standards \citep{gilmore12, randich13}. 
In this paper, we present the {\it{Gaia}}-ESO Survey selection function only for the Milky Way field stars observed with the GIRAFFE and UVES spectrographs at VLT, not including the bulge.  
All targets were selected according to their colours and magnitudes, using  photometry from the VISTA Hemisphere Survey \citep[VHS,][]{mcmahon13} and the Two Micron All-Sky Survey \citep[2MASS,][]{skrutskie06}. Selected potential target lists were generated at the Cambridge CASU centre.

We discuss the initial GIRAFFE and UVES target selection and photometry used in more detail in Sections~\ref{sec:Giraffe} and~\ref{sec:UVES}.
In the following section, we present the basic scheme constructed to select Milky Way field targets.

\subsection{Basic target selection scheme}\label{subsec:basic} 

The primary goal of the selection strategy of the survey is to select Milky Way stars in order to study a robust sample of all major Galactic components (i.e., thin and thick discs, and halo). 
The basic target selection is built on stellar magnitudes and colours. The targets are selected to sample the main sequence, the turn-off, and the red giant branch
stars centred on the red clump. To achieve this the stars are selected from two boxes, the blue and the red
(see Fig.~\ref{fig-main_boxes}a). 
The blue box is used for the selection of the turn-off and main sequence targets to be observed with GIRAFFE. The red box is defined to select stars on the red clump or nearby the red clump in the CMD. 
For the selection of stars to be observed with UVES only one box is used (see Fig.~\ref{fig-main_boxes}b).

The main GIRAFFE target selection is as follows:
\begin{equation}
\begin{split}
{\rm Blue~box:} \\
& 0.00 \le (J-K_{S})  \le 0.45;\\
& 14.0\le J \le 17.5. 
\end{split}
\label{form_G_blue_box}
\end{equation}
\begin{equation}
\begin{split}
{\rm Red~box:} \\
&  0.40 \le (J-K_{S}) \le 0.70;\\
& 12.5 \le J \le 15.0.
\end{split}
\label{form_G_red_box}
\end{equation}

The main UVES target selection is as follows:
\begin{equation}
\begin{split}
{\rm Blue~box:} \\
& 0.23 \le (J-K_{S}) \le 0.45;\\
& 12.0 \le J \le14.0.
\end{split}
\label{form_U_blue_box}
\end{equation}
Where $J$, $K_{S}$ magnitudes in Eq.~\ref{form_G_blue_box}, \ref{form_G_red_box}  are from VISTA VHS photometry and in Eq.~\ref{form_U_blue_box} they are from 2MASS photometry.
The colour boxes will be corrected for reddening, as described in Section~\ref{subsec:Reddening}.
The selection algorithm is configured to assign approximately 80$\%$ of the targets to the blue box and $\sim$ 20$\%$ to the red box for GIRAFFE (see Fig.~\ref{fig-main_boxes}a for the GIRAFFE observations). 
All the targets that are in the UVES selection box area shown in Fig.~\ref{fig-main_boxes}b were assigned as potential targets for UVES. 

\subsection{Actual target selection schemes}\label{subsec:actual} 

\begin{figure*}
\centering
\includegraphics[width=1.3\columnwidth]{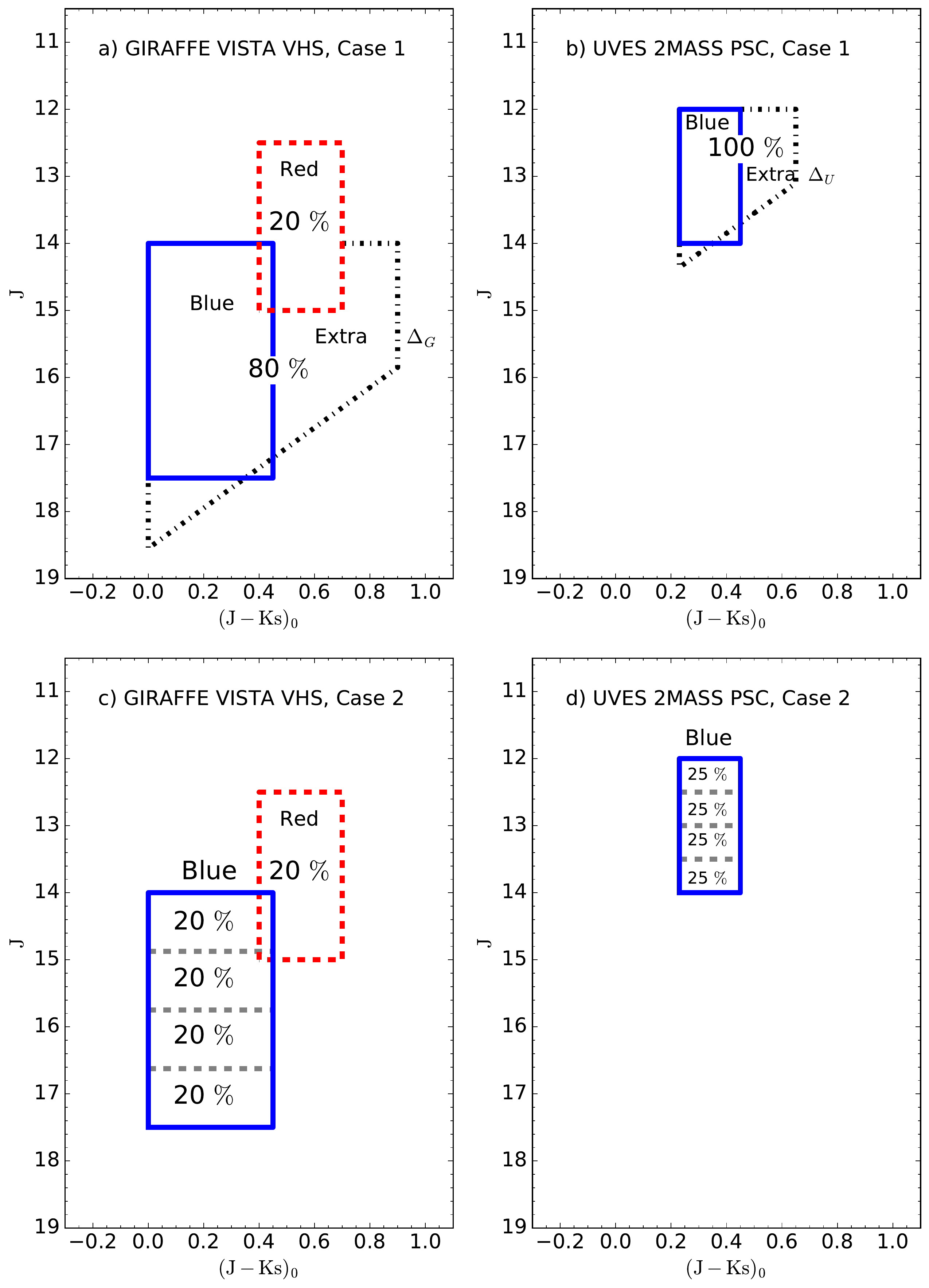}
\caption{The actual target selection colour-magnitude schemes: (a) and (c) for GIRAFFE, and (b) and (d) for UVES. Target selection based on Case 1 are shown in  (a) and (b), and on Case 2 in (c) and (d), respectively. Blue solid line shows the area of targets assigned to the blue box; red dashed line --  to the red box; and black dash-dotted line shows the area of second priority targets assigned to the extra box. The right-edge limit (a) $\Delta{_G}$ and (b) $\Delta{_U}$ in Case 1 of the extra box varies from field to field. The blue box in (c) and (d) for Case 2 is divided into 4 equal sized magnitude bins (in order to have the same number of targets per magnitude bin). On the x-axis we show the de-reddened (J-K$_{s}$)$_0$ colour, whereas on the y-axis we show the observed $J$ magnitude (for more details see Section~\ref{subsec:Reddening}).
}
\label{fig_G_U_cas1_case2}
\end{figure*}
 
We divide the selection of stars in the Milky Way fields into two cases, which are described in the following sections.

\subsubsection{Case 1}\label{subsec:case1} 

Case 1, which depends on the stellar density, occurs when the field does not have enough targets to fill the FLAMES fibres (e.g., high latitude Milky Way fields). Figures~\ref{fig_G_U_cas1_case2}a and~b show the actual target selection colour-magnitude schemes for Case~1 for GIRAFFE and UVES, respectively. 
The target selection algorithm is then extended at the right-edge of the blue boxes (see black dash-dotted box in Figs.~\ref{fig_G_U_cas1_case2}a and~b, and Eq.~\ref{form_G_extra_box}-\ref{form_U_extra_box}), 
allowing for the selection of second priority targets. 
We select second priority targets in the extra box only when all targets in the blue box have already been selected.  
In addition, in this case the second priority objects were selected with a colour-dependent $J$ magnitude
cut to avoid too faint targets, which would lead to low signal-to-noise ratios (S/N) in the optical spectra. This allows us to also extend the box to slightly fainter
magnitudes (see Figs.~\ref{fig_G_U_cas1_case2}a and b).

The second priority target selection in Case 1 for GIRAFFE is as follows:
\begin{equation}
\begin{split}
{\rm Extra~box:} \\
& 0.00 \le \big(J-K_{S}\big) \le 0.45 + \Delta_{G}; \\
& J \ge 14.0; \\
& J + 3.0*\big(\big(J-K_{S}\big)-0.35\big) \le 17.50.
\end{split}
\label{form_G_extra_box}
\end{equation}

And for UVES:
\begin{equation}
\begin{split}
{\rm Extra~box:} \\
& 0.23 \le (J-K_{S}) \le 0.45 + \Delta_{U}; \\
& J \ge 12.0; \\
& J+3.0*((J-K_{S})-0.35) \le14.00.
\end{split}
\label{form_U_extra_box}
\end{equation}

\begin{figure}
\centering
\includegraphics[width=0.8\columnwidth]{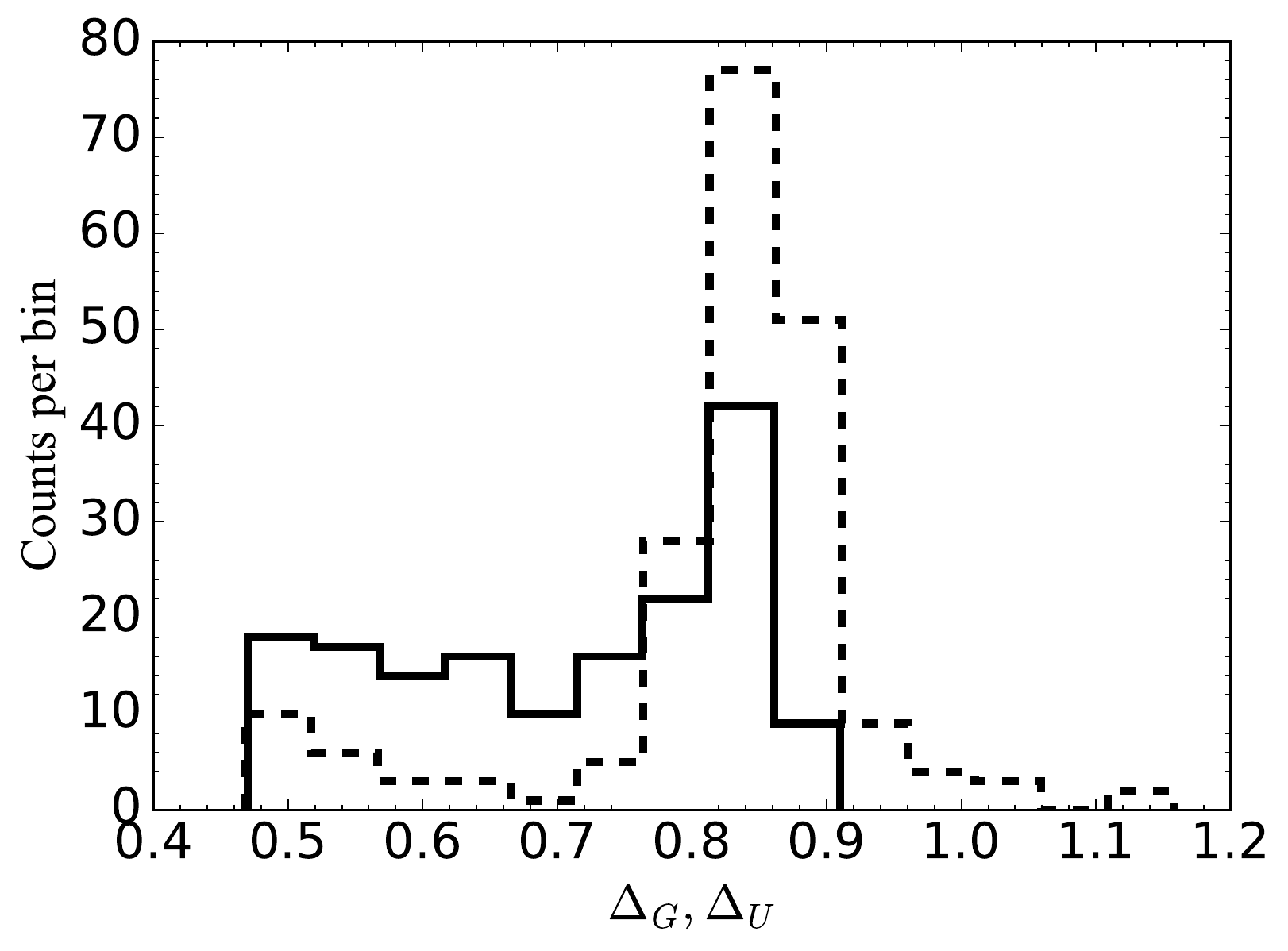}
\caption{The frequency distribution of extensions $\Delta_{G}$ and $\Delta_{U}$ (see Section~\ref{subsec:case1}). The dashed line show the right-edge extensions $\Delta_{G}$ for GIRAFFE and solid line show the right-edge extensions $\Delta_{U}$ for UVES in Case~1 Milky Way fields.
}
\label{fig_delta_g_u}
\end{figure}

Here $\Delta_{G}$ and $\Delta_{U}$ are the right-edge extensions of the extra box for GIRAFFE and UVES, respectively (see black dash-dotted line in Figs~\ref{fig_G_U_cas1_case2}a and b). Furthermore, these extensions vary slightly from field to field. The values for each field are shown in Fig.~\ref{fig_delta_g_u} and  listed in Table~\ref{table:online_T1}. 

\begin{table*}
\centering
  \caption{Main parameters and weights of the targeted and allocated Milky Way fields. The full table is available online.}
  \renewcommand{\tabcolsep}{4.6pt}
  \begin{tabular}{@{} *{32}{c} @{}}
    \hline
    GES$\_$FLD (1) & RA [h:m:S] (2) & Dec [d:m:s] (3) & E(B-V) (4) & $\Delta_{G}$ (5) & $\Delta_{U}$ (6) & W$\_$T,F$\_$b$\_$G (7) & ... $^a$ &Blue(\%) (33)\\
   \hline
    \hline
  GES\_MW\_000000-595959 & 00:00:00.000 & -59:59:59.99 & 0.012 & 0.834 & 0.834 & 0.225 & ... &  0.800\\
  GES\_MW\_000024-550000 & 00:00:24.000 & -55:00:00.00 & 0.013 & 1.033 & 0.613 & 0.272 & ... & 0.800\\
  GES\_MW\_000400-010000 & 00:04:00.000 & -01:00:00.00 & 0.034 & 0.923 & 0.763 & 0.275 & ...& 0.800\\
  GES\_MW\_000400-370000 & 00:04:00.000 & -37:00:00.00 & 0.010 & 0.855 & 0.785 & 0.237 & ...& 0.800\\
  GES\_MW\_000400-470000 & 00:04:00.000 & -47:00:00.00 & 0.009 & 0.905 & 0.636 & 0.209 & ...& 0.800\\
    ... & ... & ... & ... & ... & ... & ... & ... & ...  \\
    \hline
    \multicolumn{3}{l}{$^a$ For column names and description see Table~\ref{table:online_log}.}\\
  \end{tabular}
  \label{table:online_T1}
 \end{table*} 

\begin{figure*}
\resizebox{\hsize}{!}{
    \includegraphics[width=1\columnwidth]{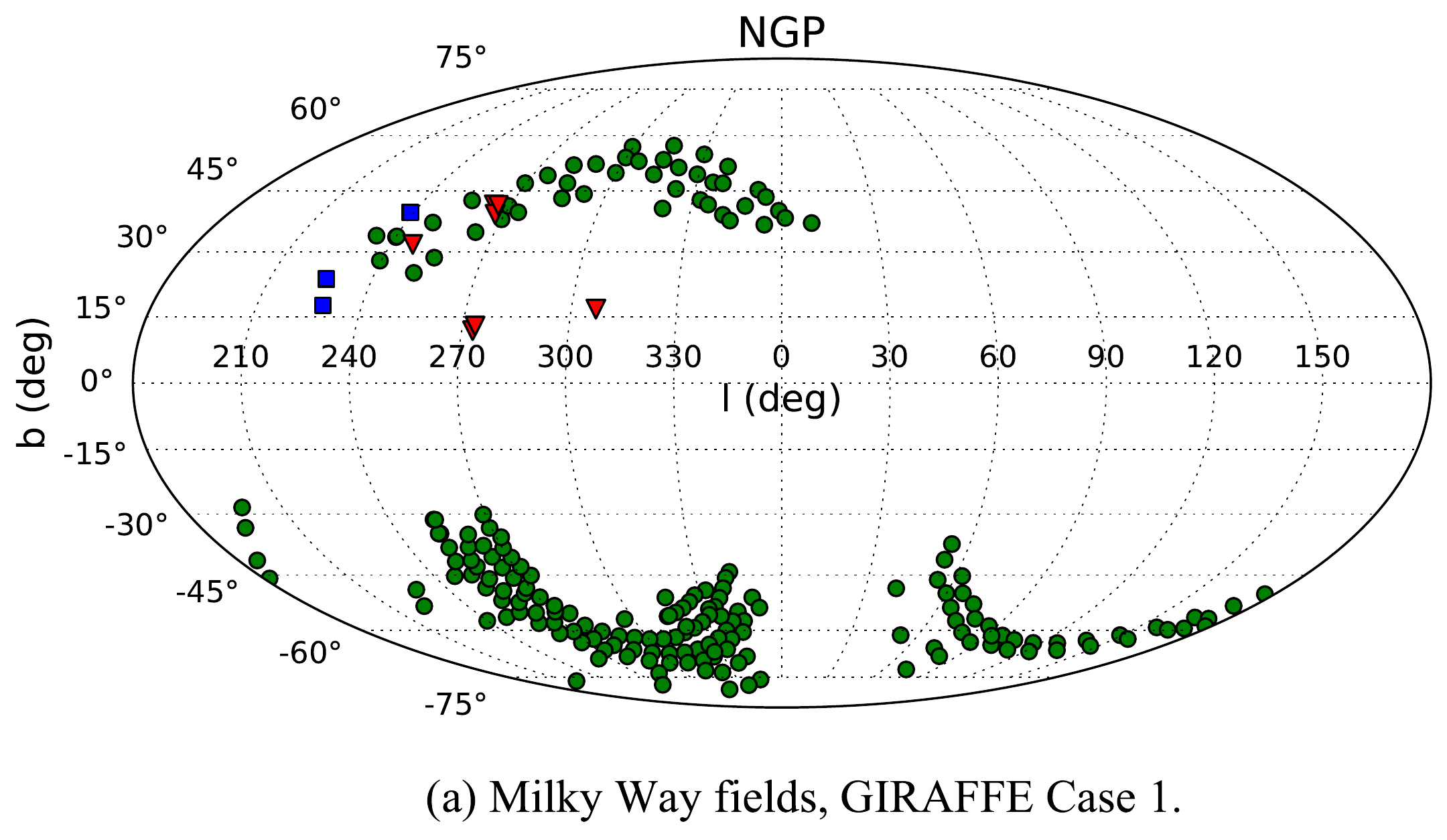}
    \includegraphics[width=1\columnwidth]{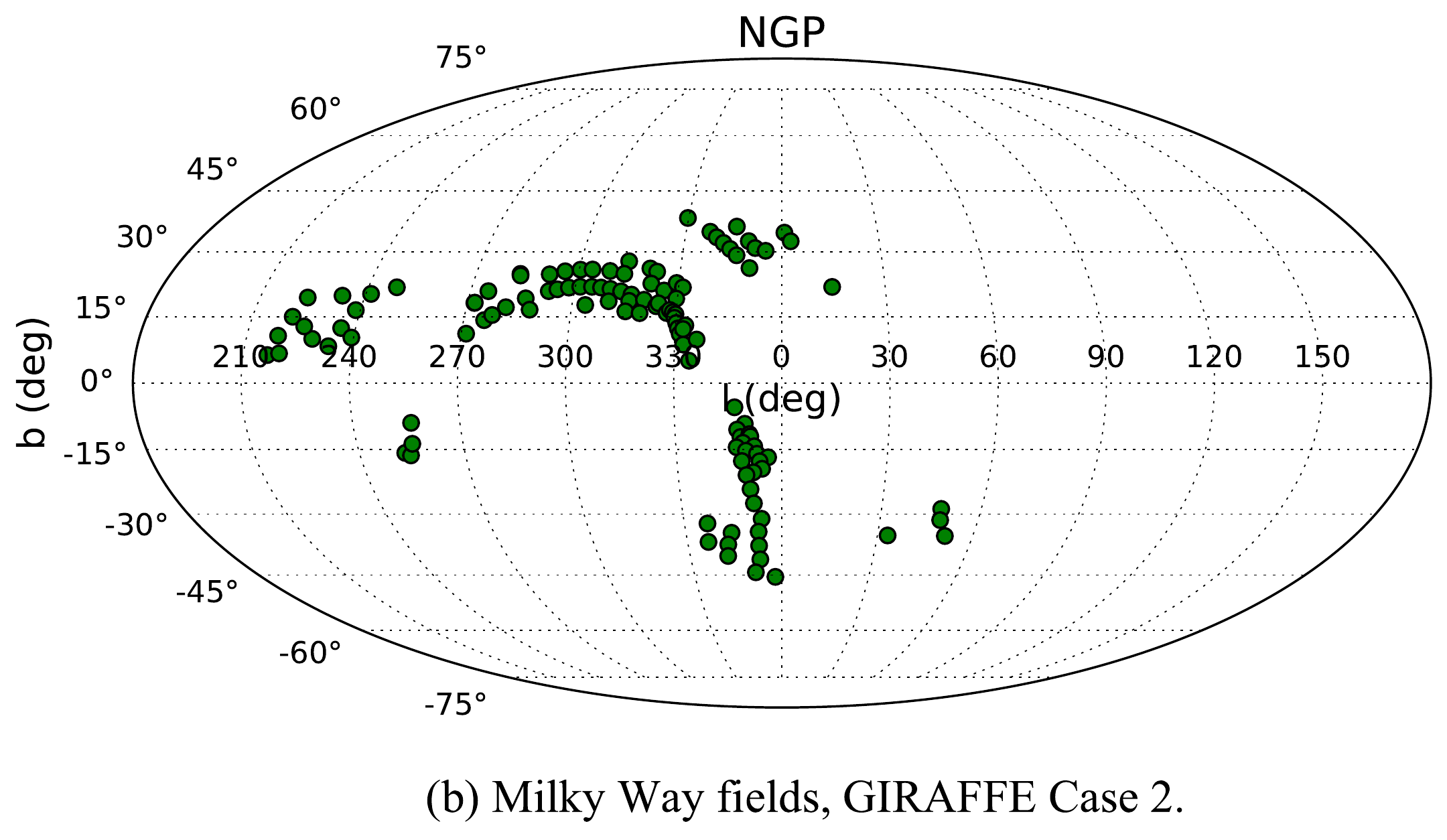}
}

\resizebox{\hsize}{!}{
    \includegraphics[width=1.\columnwidth]{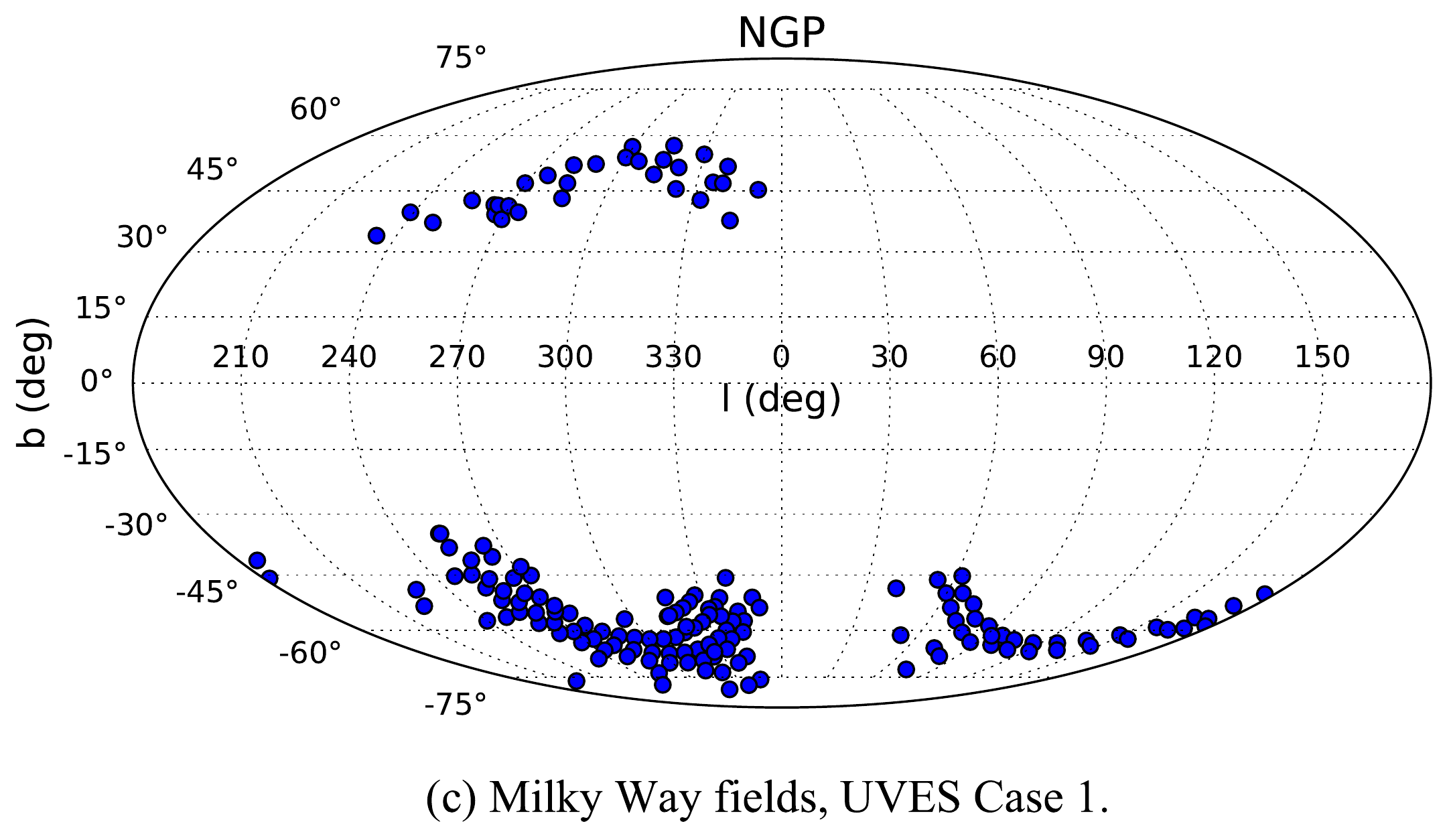}
    \includegraphics[width=1.\columnwidth]{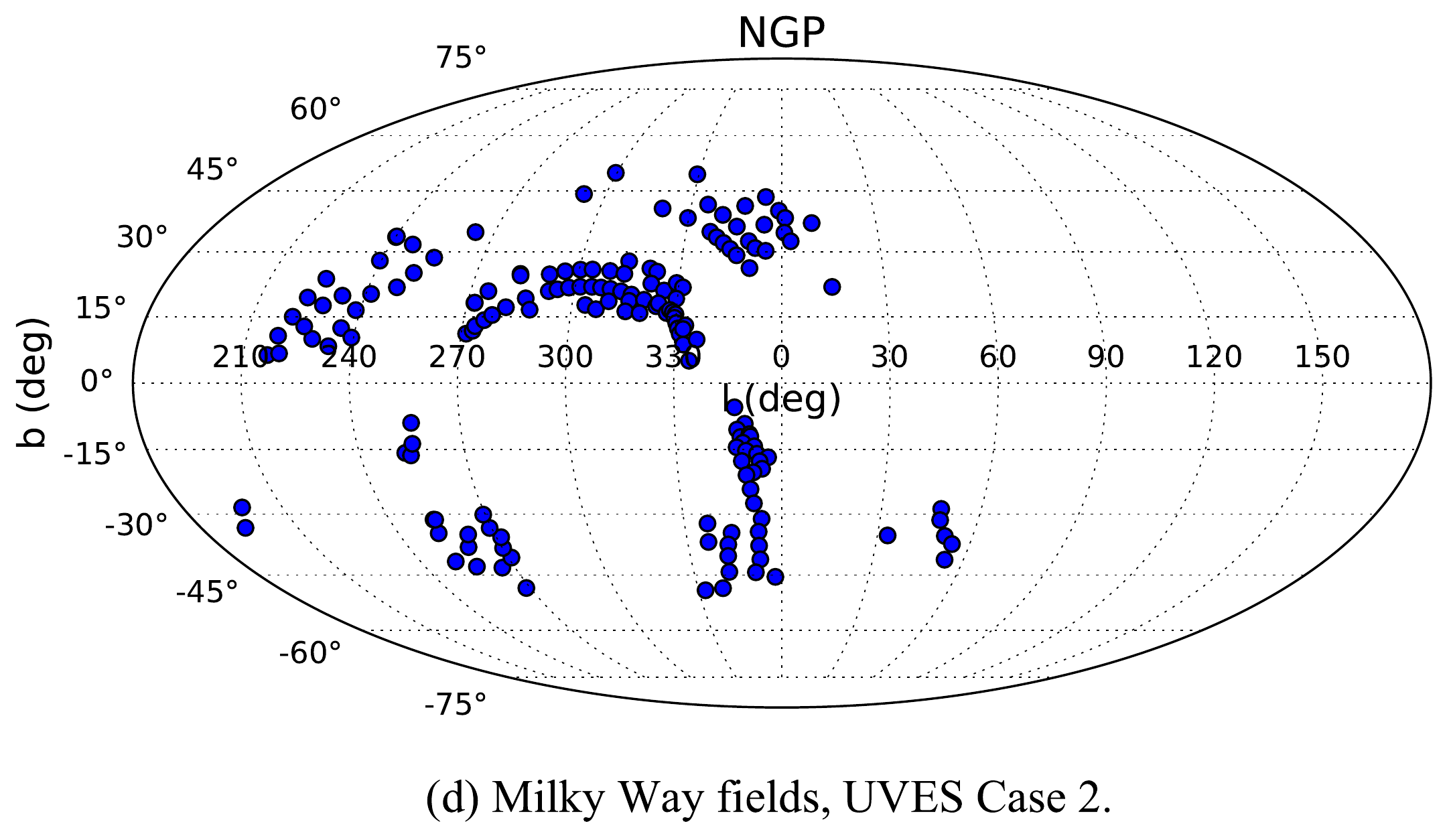}
}
\caption[]{The distribution, shown in Mollweide projection with the Galactic center in the middle, of the 
observed Milky Way fields across the sky. (a) and (b) show fields selected based on Case 1 and 2 respectively, and observed with GIRAFFE. (c) and (d) show fields selected based on Case 1 and 2 respectively, and observed with UVES. Green circles -- fields with the selection based on VISTA~VHS photometry; blue circles -- fields with the selection based on 2MASS photometry. Additional fields: blue squares --  fields with the selection based on SDSS photometry and 2MASS photometry; and red triangles -- fields with the selection based on SkyMapper photometry, VISTA VHS photometry and 2MASS photometry (for more information see Appendix~\ref{sec:AditionaFields}).}
\label{fig:Mollweide}
\end{figure*}

\begin{table*}
 \caption{The number of Milky Way fields observed in the {\it{Gaia}}-ESO Survey up to June, 2015 and in iDR4.}
 \begin{tabular}{lccccccccccccc}
  \hline
 Instrument & VHS Case~$1$  &  VHS Case~$2$ & $2$MASS Case~$1$  &  $2$MASS Case~$2$ & SDSS$^a$  &  SkyMapper$^a$ & Total  \\
  \hline
    \hline
GIRAFFE & 202 & 118 & ... & ...& 3 & 7 & 330 \\
UVES & ... &...& 164 & 166 & ... & ...& 330 \\
    \hline
iDR4 GIRAFFE & 158 & 90 & ... & ...& 2 & 7 & 257 \\
iDR4 UVES & ... &...& 128 & 135 & ... & ...& 263 \\
  \hline
 \end{tabular}
   \begin{tabular}{c}
{\it{Note.}} $^a$SDSS photometry and SkyMapper photometry were used to select additional targets (See Appendix~\ref{sec:AditionaFields}). 
 \end{tabular}
 \label{tab:no_of_fields}
\end{table*}
 
\subsubsection{Case 2}\label{subsec:case2} 

Case~2 is encountered when the density of stars exceeds the number of fibres available. 
This algorithm is applied to the Milky Way fields near to the Galactic plane.
In Case~2 the target selection algorithm selects targets in such way as to have the same number of targets per magnitude bin (i.e. not to have a bias towards very faint stars). 
Therefore, the blue box is divided into four equal-sized magnitude bins, with $J_{1,2,3,4}$=$(J_{max}-J_{min})/4$, where $J_{1}$ is the bright limit, and $J_{4}$ is the faint limit of the $J$ magnitude (see Figs.~\ref{fig_G_U_cas1_case2}c and d). In this case, we have no priorities for what to select within each given sub-box, so the choice is approximately random. 
We select approximately the same number of stars in each sub-box.
The target selection magnitudes and colours for Case~2 follow the same selection limits as presented for the main target selection (see Eq.~\ref{form_G_blue_box}-\ref{form_U_blue_box}).

\subsubsection{Extinction and colour range}\label{subsec:Reddening} 

The Milky Way fields located near the Galactic plane often suffer from considerable interstellar extinction.
The {\it{Gaia}}-ESO Survey target selection algorithm takes the line-of-sight interstellar extinction, A$_{V}$, into account using the Schlegel dust map \citep{schlegel98}. \citeauthor{schlegel98} indicate an accuracy of 16~\% on their map.
Although in the near-infrared the impact of extinction, while expected to be low, cannot be neglected, i.e. $E(J-K_{S})$~=~(c.J$-$c.K$_{s}$)$*$A$_{V}$~=~0.17$*$A$_{V}$, which leads to approximate $E(J-K_{S})$~=~0.10 for $E(B-V)$~=~0.20 \citep{rieke85}.
Here c.$J$ and c.$K_{\rm S}$ designate the extinction coefficients in the $J$ and K$_{S}$ bands (c.$J$ = A$_{J}$/A$_{V}$  and c.$K_{\rm S}$~=~A$_{K_s}$/A$_{V}$) \citep{nishiyama09}. 

The line-of-sight interstellar extinction for the GIRAFFE and UVES fields was treated differently. The line-of-sight interstellar extinction was taken into account by shifting the colour-boxes of GIRAFFE targets by 0.5$*$E$(B-V)$. Whereas for UVES targets instead the box was extended to the right (i.e. the blue edge stays fixed) by 0.5$*$E$(B-V)$. No shift was applied to the GIRAFFE and UVES boxes in the vertical, magnitude, direction. Here, E($J-K_{S}$)/E$(B-V)$~=~0.5 \citep{rieke85} and we take E$(B-V)$ as the median reddening in the field measured from \citet{schlegel98} maps. 

The median extinction estimated in the field was added to the colour boxes for GIRAFFE in the following way:

\begin{equation}
\begin{split}
&{\rm Blue~box:} \\
& 0.5E(B-V)+[0.00 \le (J-K_{S}) \le 0.45];\\
&{\rm Red~box:} \\
& 0.5E(B-V)+[0.40 \le (J-K_{S}) \le 0.70];\\
&{\rm Extra~box:} \\
& 0.5E(B-V)+[0.00 \le (J-K_{S}) \le 0.45+\Delta_{G}].\\
\end{split}
\label{form_G_blue_box_ext}
\end{equation}

And for UVES:
\begin{equation}
\begin{split}
&{\rm Blue~box:} \\
&  0.23 \le (J-K_{S}) \le 0.45+0.5E(B-V);\\
&{\rm Extra~box:} \\
&  0.23 \le (J-K_{S}) \le 0.45+\Delta_{U}+0.5E(B-V).\\
\end{split}
\label{form_U_rblue_box_ext}
\end{equation}

The median  E$(B-V)$ values vary per field and are listed in Table~\ref{table:online_T1}.
The mean of the line-of-sight reddening for the {\it{Gaia}}-ESO Survey observed in Case~1 Milky Way field stars never reaches values greater than E$(B-V)$~=~0.10, and for fields near the Galactic plane not greater than E$(B-V)$~=~1.23. The mean line-of-sight reddening value for Case~1 fields is <E$(B-V)$>~=~0.03$\pm$0.02. For Case~2 Milky Way fields located near the Galactic plane the mean line-of-sight reddening value is <E$(B-V)$>~=~0.10$\pm$0.12.

\begin{table*}
 \caption{Adopted data quality flags for VHS photometry (GIRAFFE) and 2MASS photometry (UVES ), and SDSS$^a$ photometry.}
 \begin{tabular}{lllcccccccc}
  \hline
Catalog& Requirement & Notes\\
  \hline
    \hline
VHS mergedClass						   & $mergedClass=-1$								&Classified as a star\\
VHS jAverageConf, ksAverageConf			   & $ jAverageConf~$>$~95,~ksAverageConf~$>$~95$		&Average confidence in $J$, $K_{\rm S}$ mag\\
VHS jErrBits, ksErrBits					   & $jErrBits~$=$~0,~ksErrBits~$=$~0$					&Warning/error bitwise flags in $J$, $K_{\rm S}$ mag\\
VHS Not on the bad $CCD$				   &  $jx$<$8800~OR~jy$<$12300	$					&Flags used in internal release\\
2MASS ph\_qual						   &  $ph\_qual=AAA$								&Photometric quality flag \\
SDSS mode					   		   &  $mode=1$										&Flag indicates primary sources \\
SDSS gc						   		   &  $type=6$										&Phototype in $g$ band, 6=Star \\
  \hline
 \end{tabular}
  \begin{tabular}{c}
{\it{Note.}} $^a$SDSS photometry was used to select additional targets (See Appendix~\ref{sec:sdss}). 
 \end{tabular}
 \label{tab:flags}
\end{table*}

\subsubsection{Naming conventions}\label{subsec:names} 

The {\it{Gaia}}-ESO Survey Milky Way field names were created at Cambridge Astronomy Survey Unit (CASU) from the right ascension ``hms'' and declination ``dms'' (J2000) of the field center. For example, the {\it{Gaia}}-ESO Survey Milky Way field centered at RA$=$14$^{h}$20$^{m}$00$^{s}$ and Dec$=$$-$05$^{d}$00$^{m}$00$^{s}$ was assigned the name GES$\_$MW$\_$142000-050000. 
The names of the selected targets (objects)  also encode which selection criteria were used to select them. ``$\_b\_$'' means the blue box, ``$\_r\_$''  the red box, and ``$\_e\_$''  is for the extra box (identifies the objects which were added to fill the fibres).
Some of the targets were selected by both blue and red boxes, and they have ``$\_br\_$''  in their name.

In the fits headers of the data files the Milky Way fields can be identified with the keyword ``$GES\_TYPE$''.
For the Milky Way fields it is set to ``$GE\_MW$''.


\subsubsection{The Milky Way field pattern}\label{sec:MWpatern}

The distribution of observed Milky Way fields in the {\it{Gaia}}-ESO Survey is designed to be well spread.
However, the observation range in the Galaxy is restricted to $+$10$^\circ$~$\ge$~Dec~$\ge$~$-$60$^\circ$ to minimise the airmass. Figure~\ref{fig:Mollweide} shows the distribution of so far observed fields on the sky.
Table~\ref{tab:no_of_fields} lists the number of observed Milky Way fields in the {\it{Gaia}}-ESO Survey up to June 2015 and the number of Milky Way fields in iDR4. For these fields targets were selected as outlined in the preceding sections.

A small subset of additional Milky Way field targets were selected using SDSS and SkyMapper photometry in order to study metal-poor stars and K~giants. 
Some details on the selection of these targets are given in Appendix~\ref{sec:AditionaFields}. 

\begin{figure}
\includegraphics[width=\columnwidth]{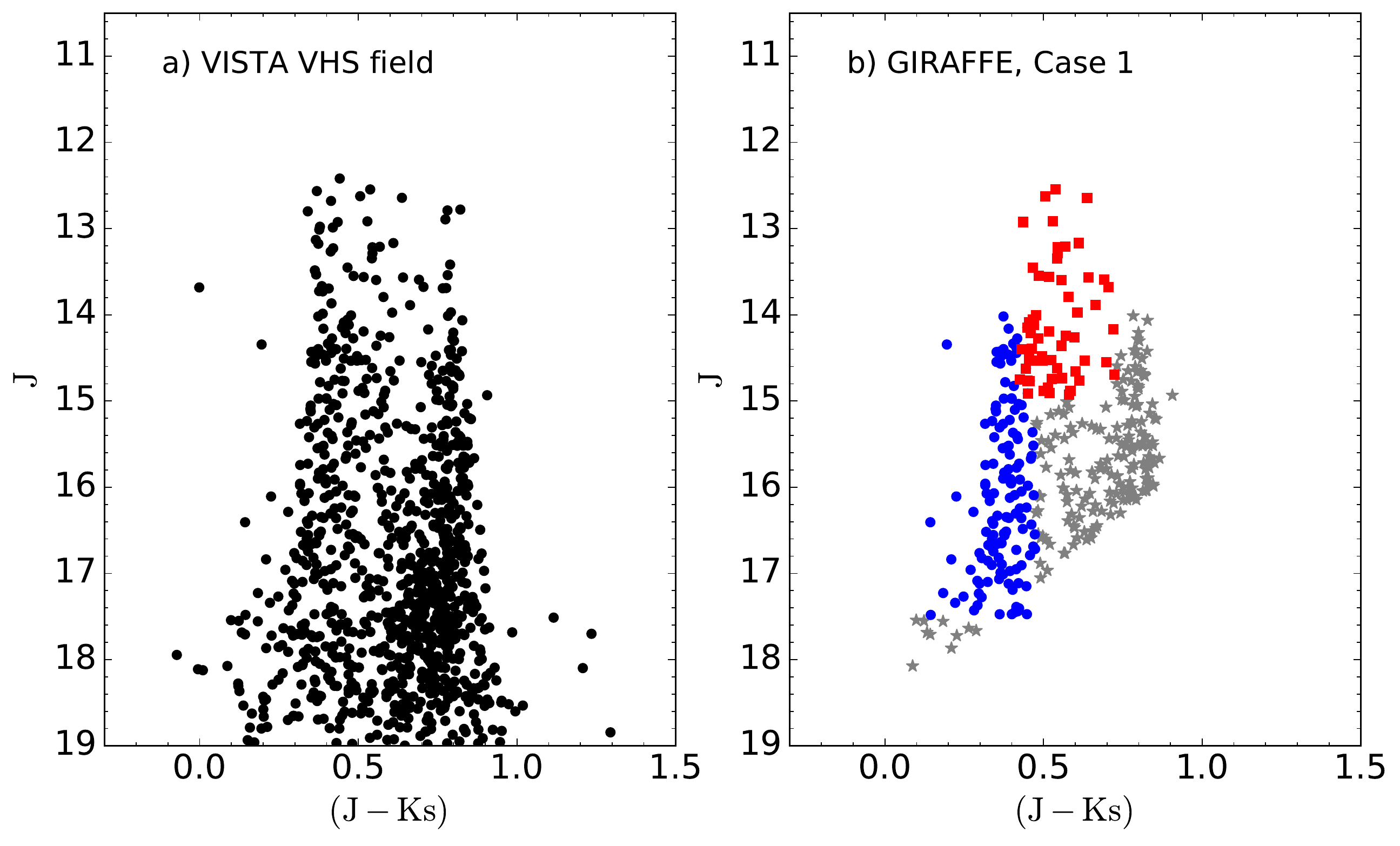}
\caption{Colour-magnitude diagrams with VISTA~VHS photometry. (a)  CMD of the field GES{\_}MW{\_}142000-050000  centered  at Galactic longitude $l=$339.9$^{\circ}$ and latitude $b=$51.4$^{\circ}$, and FoV=35$^{\prime}$ in diameter. (b) GIRAFFE target selection based on Case~1 selection scheme (see Section \ref{subsec:case1}). Blue circles -- selection of targets in blue box; red squares -- selection of targets in red box; 
and grey stars -- second priority targets, respectively. }
\label{fig-case1_g}
\vspace{0mm}
\end{figure}

\begin{figure}
\includegraphics[width=\columnwidth]{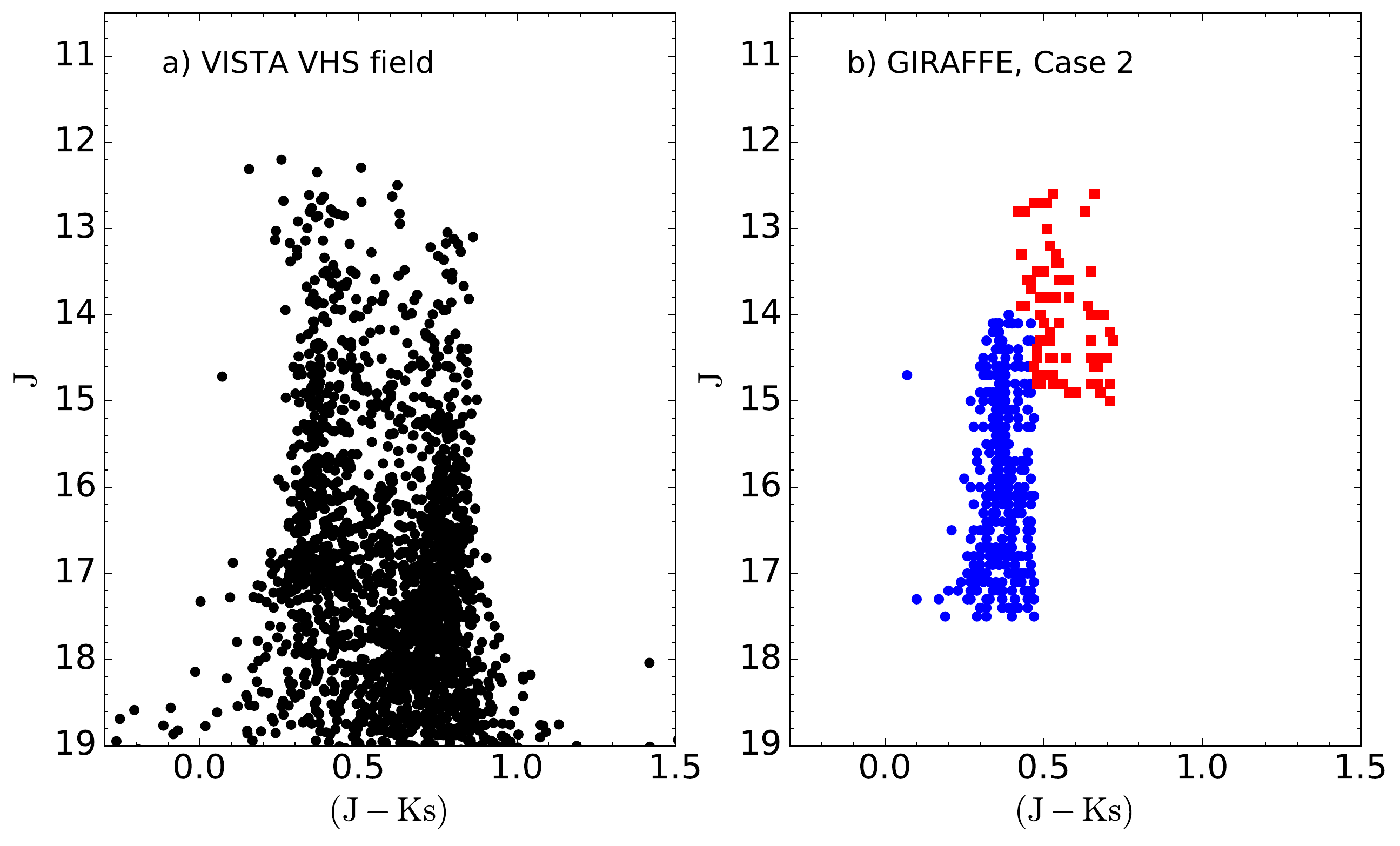}
\caption{Colour-magnitude diagrams with VISTA~VHS photometry. (a) CMD of the field GES{\_}MW{\_}201959-470000 centered at Galactic longitude  $l=$352.7$^{\circ}$ and latitude$b=$$-$34.2$^{\circ}$, and FoV=35$^{\prime}$ in diameter. (b) GIRAFFE target selection based on Case~2 selection scheme (see Section \ref{subsec:case2}). Blue circles -- selection of targets in blue box; red squares -- selection of targets in red box, respectively. }
\label{fig-case2_g}
\vspace{0mm}
\end{figure}

\section{Initial GIRAFFE target selection}\label{sec:Giraffe} 

\subsection{Photometric catalog}\label{subsec:GiraffeConstarins} 

The survey input and target selection catalogue is the VISTA Hemisphere Survey (VHS) for the Milky Way fields observed with GIRAFFE \citep{mcmahon13}. 
The target selection is based on the panoramic wide field infrared VISTA Hemisphere Survey (VHS). The VHS survey data consists of three survey components:
VHS Galactic Plane Survey (VHS-GPS); VHS-ATLAS and VHS-Dark Energy Survey (VHS-DES). In particular, catalog versions from 2011 to 2014 were used to select Milky Way field targets. 
VISTA~VHS has a sufficient sky coverage to meet the full science goals of the {\it{Gaia}}-ESO Survey. 
This catalogue is $\sim$ 30 times deeper than the Two Micron All Sky Survey (2MASS) in at least two wavebands ($J$ and $K_{S}$) \citep{mcmahon13} (see more about VISTA~VHS\footnote{http://www.vista-vhs.org}).
The adopted data quality flags to select Milky Way field targets from the VISTA Hemisphere Survey catalog are listed in Table \ref{tab:flags}.

The target selection magnitude and colour limits for GIRAFFE are presented in Sections \ref{subsec:basic} and \ref{subsec:actual}. 

\subsection{Target selection Case 1 and 2}\label{subsec:GiraffeCase1}

An example of the target selection for Case~1 is shown in Fig.~\ref{fig-case1_g}. Here, the selected blue circles are targeted turn-off and main sequence stars to be observed with GIRAFFE, and the red squares are red clump stars.
An example of the target selection for Case~2 is shown in Fig.~\ref{fig-case2_g}.
For Case~2 the target selection algorithm selected roughly the same number of blue box targets per $J$ magnitude bin.

As mentioned before, for most of the Milky Way fields the initial target selection algorithm tried to assign 80 \% of the targets to the blue box and 20 \% to the red box, respectively. The selection for some
of the fields near the Galactic bulge were the only fields where the selection of a blue versus red box fraction
was changed, i.e. from 80/20~\% to 20/80~\%, to predominantly observe star the Galactic bulge direction, i.e. the red clump stars. Those Milky Way fields are indicated in the last column of Table~\ref{table:online_T1}. 


\section{Initial UVES target selection}\label{sec:UVES} 

The {\it{Gaia}}-ESO Survey uses UVES  with the U580 setup (470-684~nm) to observe Milky Way field stars. Up to 7 separate objects (plus one sky fibre) can be allocated and observed simultaneously in the U580 mode. 
The methodology adopted in the {\it{Gaia}}-ESO Survey is such that the Milky Way field observations with UVES are made in parallel with the GIRAFFE field star observations.
This means that the exposure times are planned according to the observations being executed with the GIRAFFE fibres. The UVES targets are chosen according to their near-infrared colours to be FG-dwarfs/turn-off stars with magnitudes down to $J_{2MASS}$$=$14~mag. 

The target selection box for UVES was defined using the Two Micron All Sky Survey Point Source Catalog (2MASS PSC) photometry \citep{skrutskie06}.
VISTA VHS photometry suffers saturation in the relevant magnitude range while 2MASS delivers better photometry.

\begin{figure}
\includegraphics[width=\columnwidth]{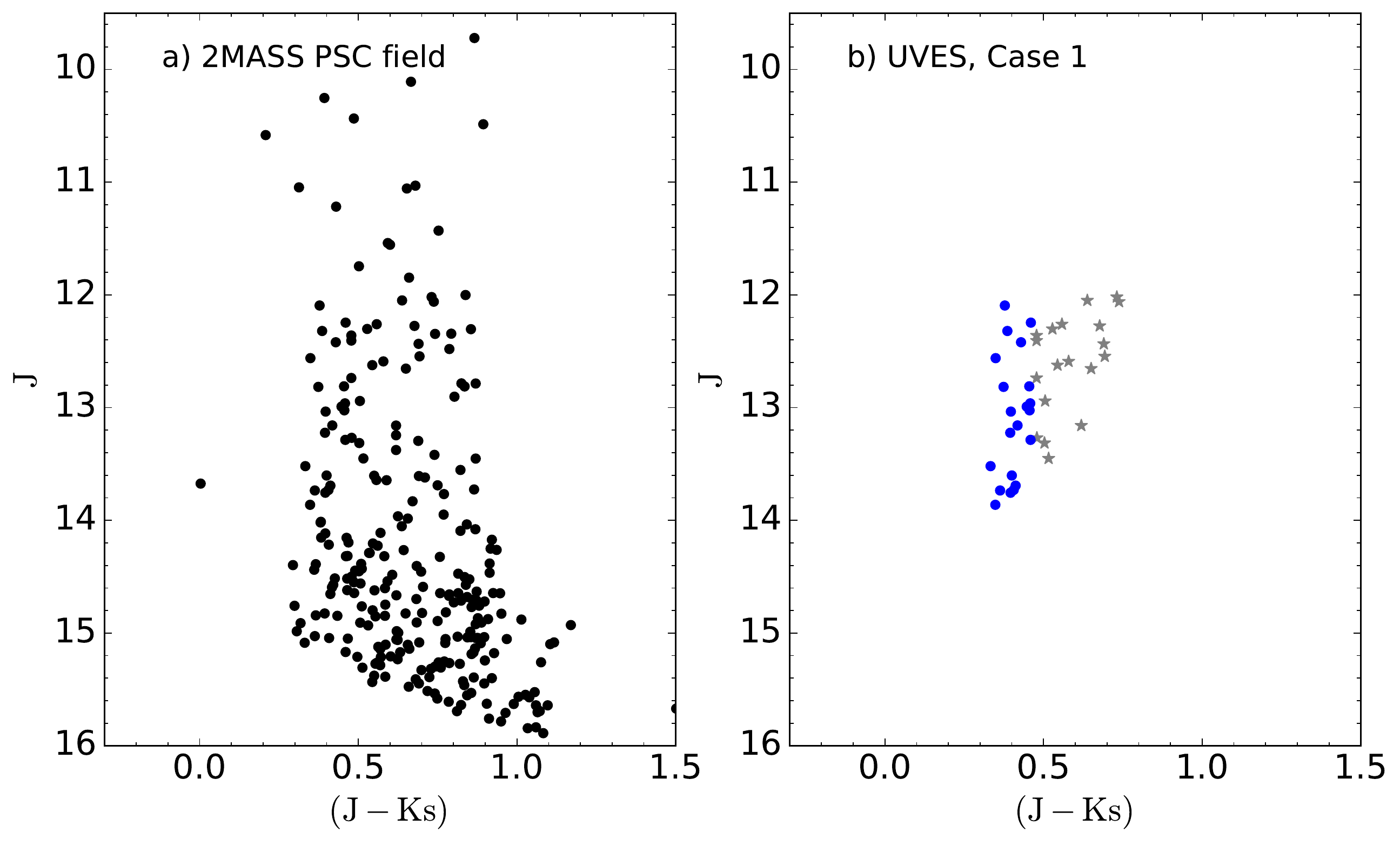}
\caption{Colour-magnitude diagrams with 2MASS~PSC photometry. (a)  CMD of the field GES{\_}MW{\_}142000-050000 centered at Galactic longitude  $l=$339.9$^{\circ}$ and  latitude $b=$51.4$^{\circ}$, and FoV=35$^{\prime}$ in diameter. (b) UVES target selection based on Case 1. Blue circles -- selection of targets in the blue box, and grey stars -- extra targets, respectively. }
\label{fig-case1_u}
\vspace{0mm}
\end{figure}

\begin{figure}
\includegraphics[width=\columnwidth]{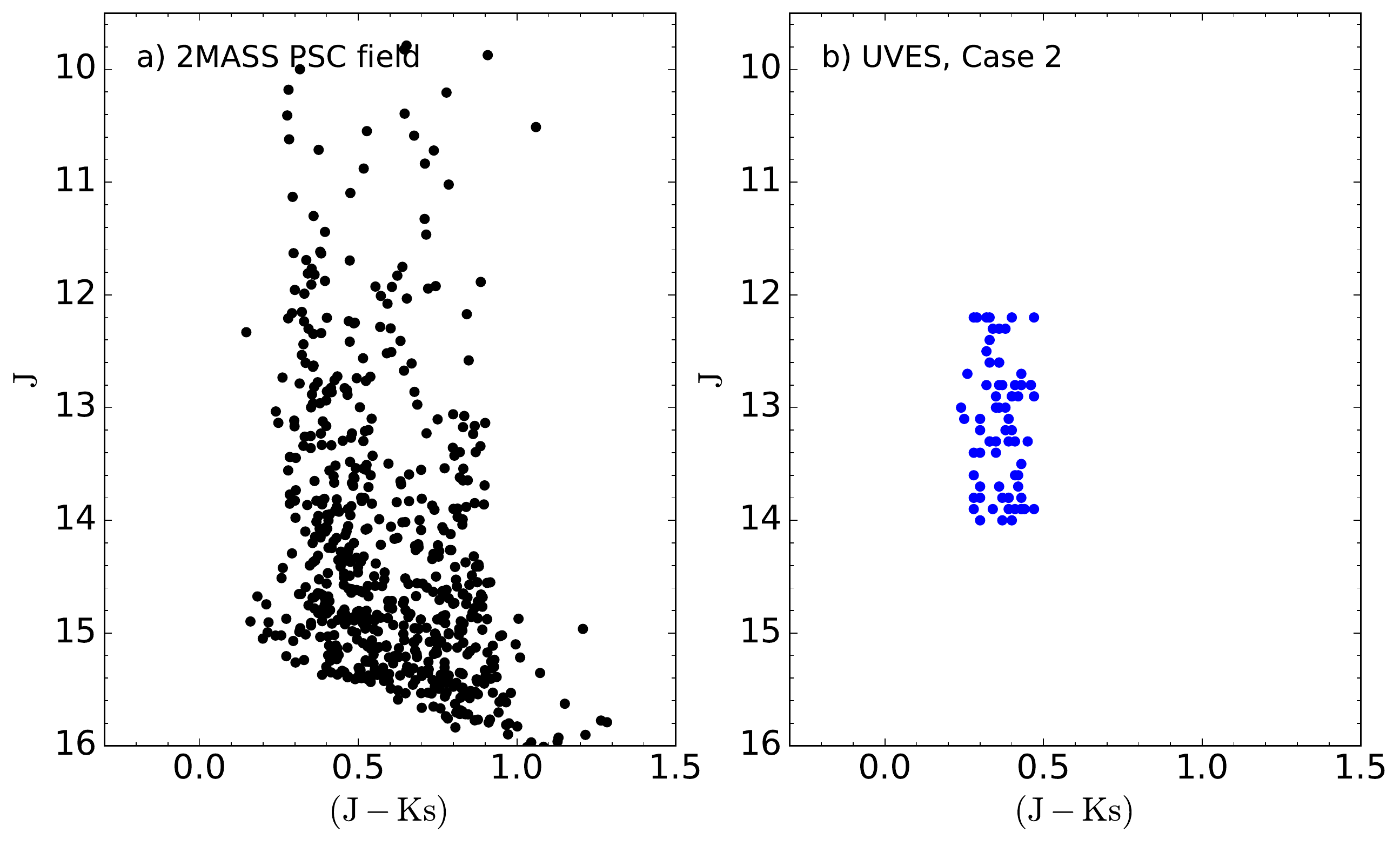}
\caption{Colour-magnitude diagrams with 2MASS~PSC photometry. (a)  CMD of the field GES{\_}MW{\_}201959-470000 centered at Galactic longitude $l=$352.7$^{\circ}$ and latitude  $b=$$-$34.2$^{\circ}$, and FoV=35$^{\prime}$ in diameter. (b) UVES target selection based on Case 2 showing the selected targets in the blue.}
\label{fig-case2_u}
\vspace{0mm}
\end{figure}

As for the case of GIRAFFE, the \citet{schlegel98} dust map, A$_{V}$, was used to determine the median extinction in the field. The 2MASS catalog flag used to select Milky Way field targets is given in Table \ref{tab:flags}.
The target selection algorithm for UVES is configured using the same methodology as for GIRAFFE and is presented in Section~\ref{subsec:basic} and \ref{subsec:actual}.   
There is only one difference in the target selection limits for UVES targets. The UVES target selection for six Milky Way fields based on Case~1 and for twenty fields for Case~2 have the brightest cut on $J_{2MASS}$ of 11~instead of 12~mag and these are listed in Table \ref{table:uves_11j}. 
The target selection maximum magnitude range for UVES in $J_{2MASS}$ is 2.0 magnitudes within the narrow range of $(J-K_{S})$.
Figures \ref{fig-case1_u} and \ref{fig-case2_u} show the UVES target selection for two Milky Way fields for Case~1 and~2, respectively.

\begin{table}
\centering
\caption{UVES Milky Way fields with $J$$=$11 mag selection limit.}
\begin{tabular}{lccl}
\hline
Case 1 \\
\hline
\hline
GES$\_$MW$\_$025559-003000\\
GES$\_$MW$\_$031800-003000\\
GES$\_$MW$\_$033800-273000\\
GES$\_$MW$\_$033959-000000\\
GES$\_$MW$\_$092800-003000\\
GES$\_$MW$\_$112200-100000\\
\hline
Case 2 \\
\hline
\hline
GES$\_$MW$\_$041959-001959\\
GES$\_$MW$\_$050000-520000\\
GES$\_$MW$\_$070359-423000\\
GES$\_$MW$\_$072048-003000\\
GES$\_$MW$\_$074500-423000\\
GES$\_$MW$\_$075600-090000\\
GES$\_$MW$\_$075959-003000\\
GES$\_$MW$\_$100000-410000\\
GES$\_$MW$\_$105959-410000\\
GES$\_$MW$\_$120000-410000\\
GES$\_$MW$\_$124224-130559\\
GES$\_$MW$\_$130047-410000\\
GES$\_$MW$\_$140000-100000\\
GES$\_$MW$\_$140000-410000\\
GES$\_$MW$\_$145800-410000\\
GES$\_$MW$\_$150159-100000\\
GES$\_$MW$\_$155400-410000\\
GES$\_$MW$\_$155959-003000\\
GES$\_$MW$\_$170024-051200\\
GES$\_$MW$\_$173359-430000\\
\hline
\end{tabular}
\label{table:uves_11j}
\end{table}


\section{Final target selection: allocating the fibres}\label{sec:Final}

The selected potential target lists were generated using the methodology presented in Sections \ref{subsec:basic} and~\ref{subsec:actual}. 
Thereafter,  the observing team generated
the final target allocation catalog that was used for the actual observations at the VLT.

\begin{figure*}
\resizebox{0.7\hsize}{!}{
\includegraphics[width=\columnwidth]{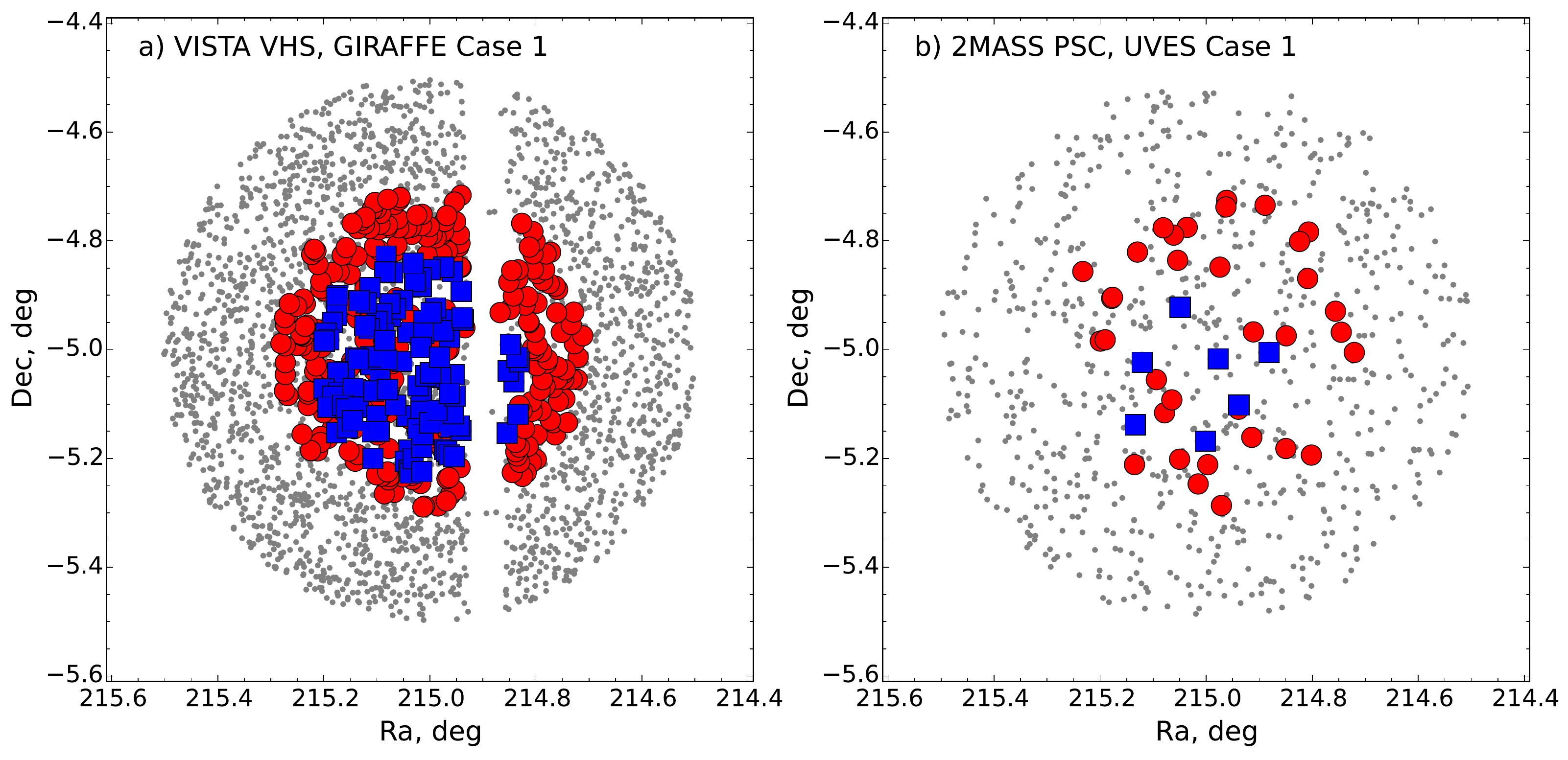}
}
\resizebox{0.7\hsize}{!}{
\includegraphics[width=\columnwidth]{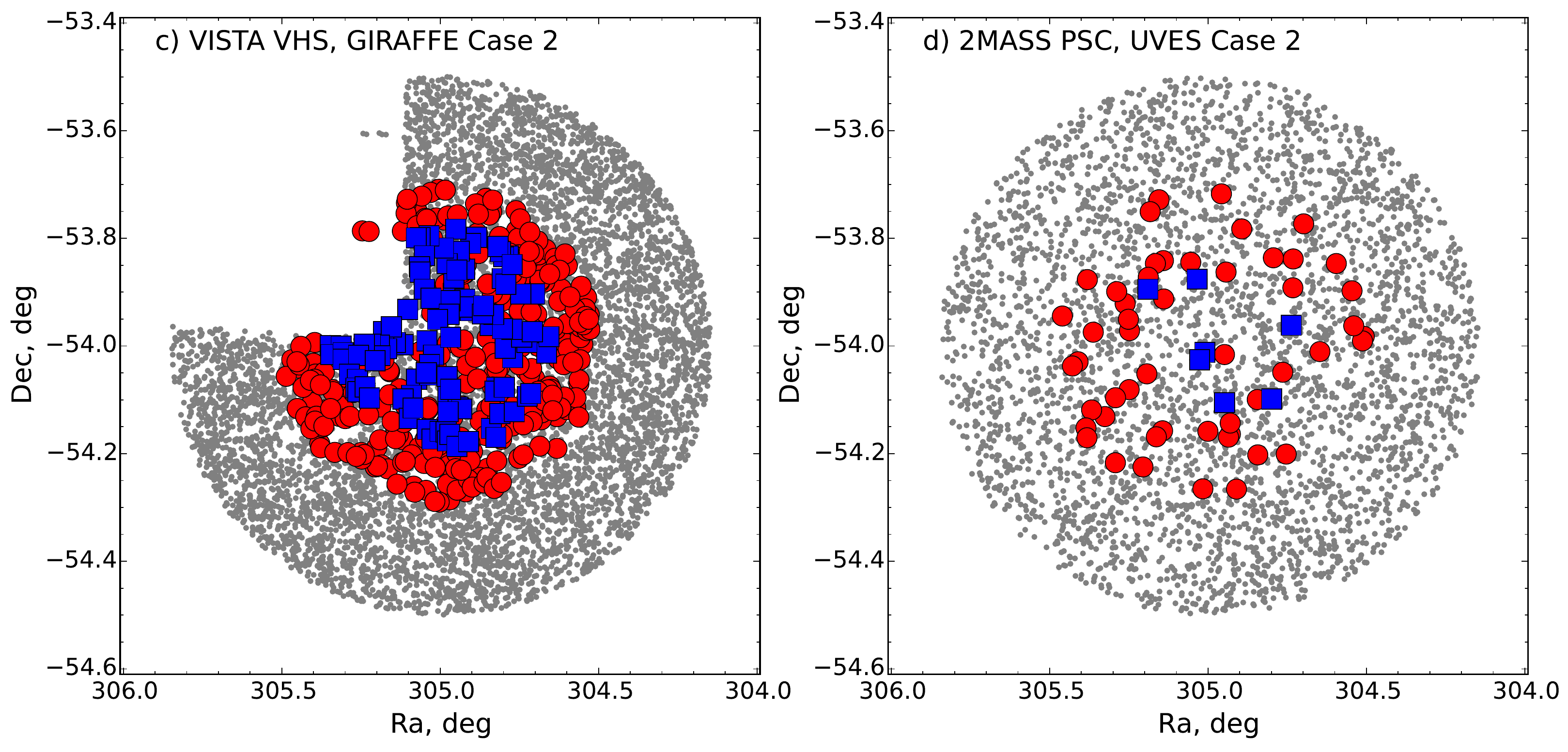}
}
\caption{Spatial target distribution on the sky. 
(a) and (b) Milky Way field GES\_MW\_142000-050000 centered  at Galactic $l$=339.9$^{\circ}$ and $b$=51.39$^{\circ}$, and selected based on Case 1 and VISTA VHS, 2MASS photometry;
and (c), (d) field GES\_MW\_201959-540000 centered  at Galactic $l$=344.3$^{\circ}$ and $b$=$-$34.5$^{\circ}$, and selected based on Case 2, respectively. 
Grey dots -- targets distributed in a 1-square-degree FoV in diameter; red filled circles -- targets within 35$^{\prime}$ FoV in diameter; blue filled squares -- allocated FLAMES targets, with 25$^{\prime}$ FoV in diameter.
}
\label{fig-allocated}
\end{figure*}

The target list has a larger FoV (35$^{\prime}$)  in diameter than the FoV for FLAMES and hence has a larger number of targets per field than can be allocated on FLAMES (see Fig.~\ref{fig-FoV} and red filled circles versus blue squares in Fig.~\ref{fig-allocated}).
This large size of the potential target list is motivated by the fact that for each observing block a guide star must also be allocated and that it is of interest to allocate as many fibres as possible. To allow for some flexibility of the center of the final field the list of potential targets hence covers a larger area on the sky. The centers of the allocated target lists are close to the original field centers.

The observing team uses the Fibre Positioner Observation Support Software (FPOSS, see the user manual\footnote{https://www.eso.org/sci/facilities/paranal/instruments\\ /flames/doc/VLT-MAN-ESO-13700-0079\_v93.pdf} for more details), which is the fibre configuration program for the preparation of FLAMES observations. FPOSS takes as input a file containing a list of target objects and generates a configuration in which as many fibres as possible are allocated to targets, allowing for the various instrumental constraints and any specified target priorities. It produces a file containing a list of allocations of fibres to targets, the so-called target setup file.

The final {\it{Gaia}}-ESO Survey target selection function depends on the allocated and observed targets. 
An illustration of two fields from Case~1 and Case~2 is shown in Fig.~\ref{fig-allocated}. Here targets are shown spatially distributed on the sky within the three different field-of-view 
introduced in Fig~\ref{fig-FoV} and discussed earlier in this section. Grey dots show targets distributed in a 1-square-degree FoV  in diameter; red filled circles show targets within 35$^{\prime}$ FoV  in diameter (the one used to make the allocations); and blue filled squares show allocated FLAMES targets, with 25$^{\prime}$ FoV in diameter for two
different Milky Way fields centered at Galactic longitude $l$$=$339.9$^{\circ}$,  latitude $b$$=$51.39$^{\circ}$ and $l$$=$344.3$^{\circ}$, $b$$=-$34.5$^{\circ}$, respectively.

There are several interesting points to extract~from~this illustration. First, it can be seen by visual inspection that Figs~\ref{fig-allocated}a and c show the incompleteness of the VISTA~VHS catalog at the time when the catalog was used for GIRAFFE target selection. Figures~\ref{fig-allocated}a and b show Case~1 for GIRAFFE and UVES respectively.
In this example a total of 111 targets (including 33 second priority targets) were allocated on FLAMES/GIRAFFE for the Milky Way field centered at Galactic  longitude $l$$=$339.9$^{\circ}$ and latitude $b$$=$51.39$^{\circ}$ (Fig.~\ref{fig-allocated}a). 
The total number of allocated targets on FLAMES/UVES for the same field is seven (including three as second priority targets) (Fig~\ref{fig-allocated}b). 
The rest of the fibres were sky fibres. 
Figures~\ref{fig-allocated}c and d show an example of Case~2, for the field GES\_MW\_201959-540000 centered  at Galactic $l$$=$344.3$^{\circ}$ and $b$$=-$34.5$^{\circ}$. A total of 104 fibres were allocated (with $\sim$ 80~\% from the blue box and $\sim$ 20~\% from the red box) for GIRAFFE and 7 for UVES. 

\begin{figure*}
\centering
\includegraphics[width=1.65\columnwidth]{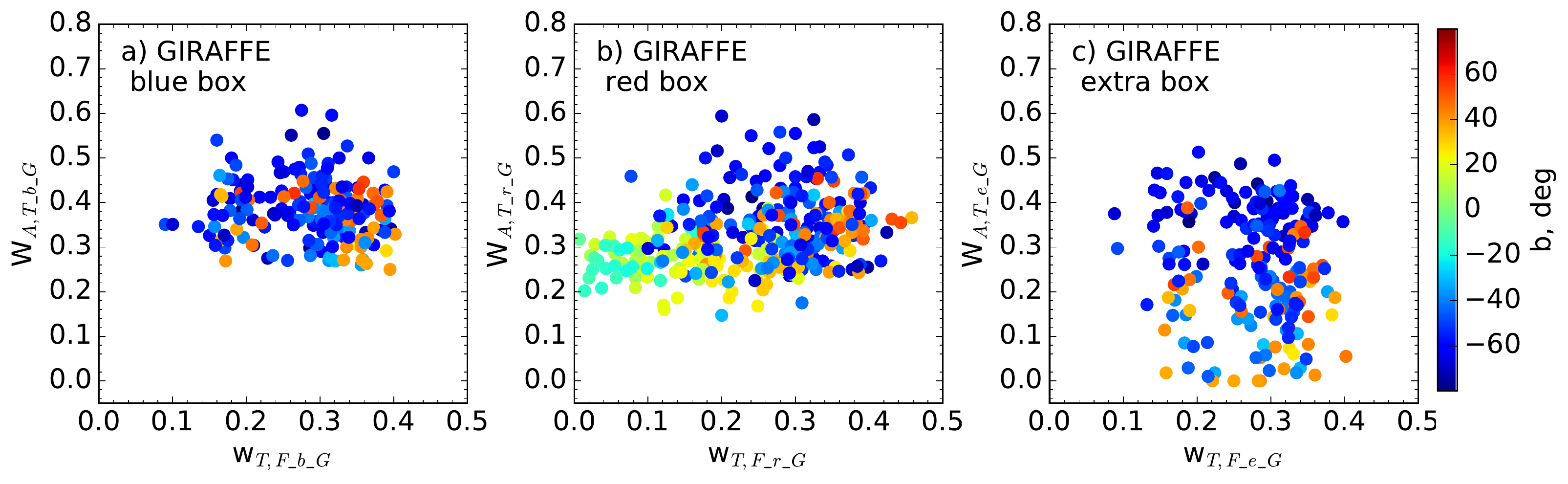}
\caption{Weights of each field for targeted stars versus stars in the field within a 1-degree FoV in diameter compared with weights for allocated versus targeted stars in (a) blue, (b) red and (c) extra boxes for GIRAFFE. The colour coding indicates Galactic latitude in degrees. 
}
\label{fig-weights1}
\end{figure*}

\begin{figure}
\includegraphics[width=1\columnwidth]{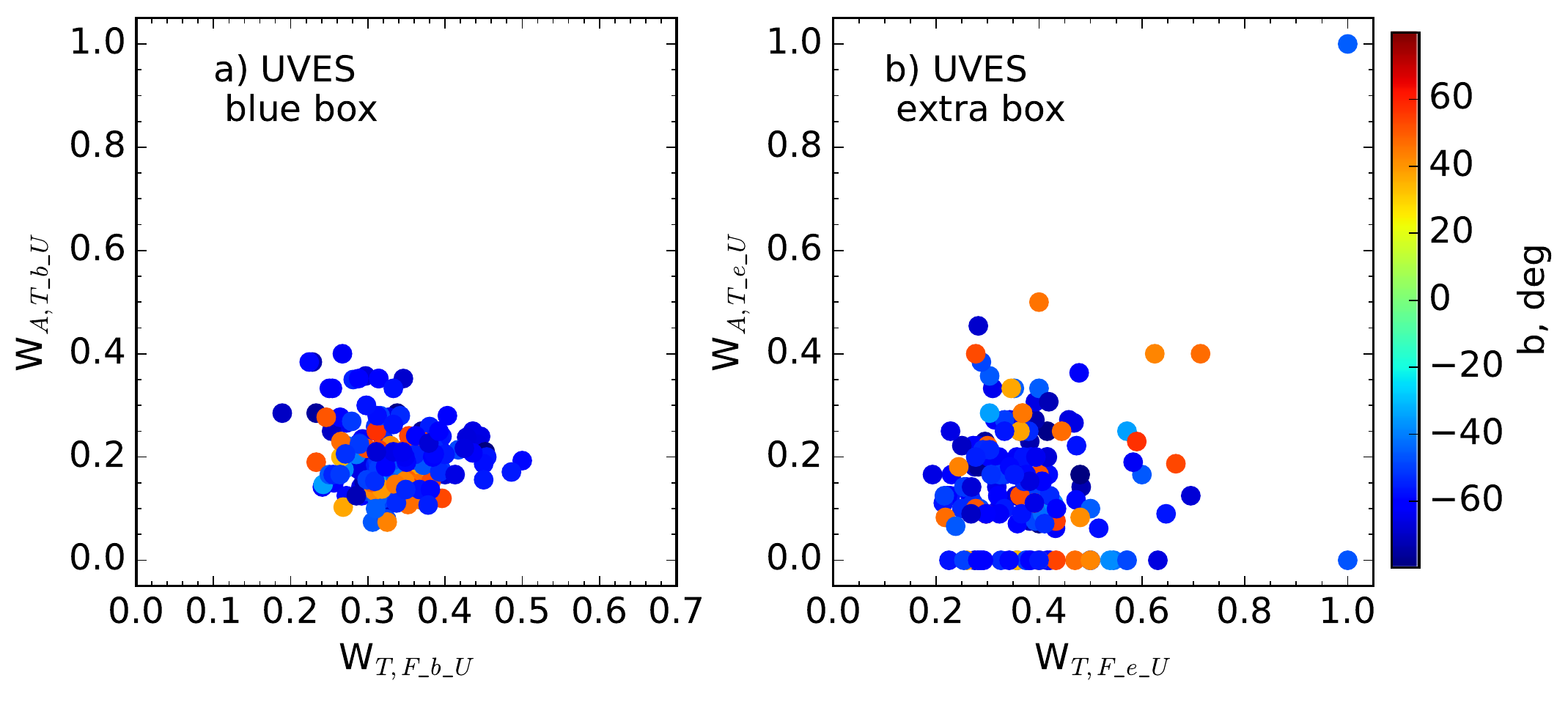}
\caption{Weights of each field for targeted stars versus stars in the field within a 1-degree FoV in diameter compared with weights for allocated versus targeted stars in (a) blue and (b) extra boxes for UVES. The colour coding indicates Galactic latitude in degrees. 
}
\label{fig-weights2}
\end{figure}

\begin{figure}
\includegraphics[width=1\columnwidth]{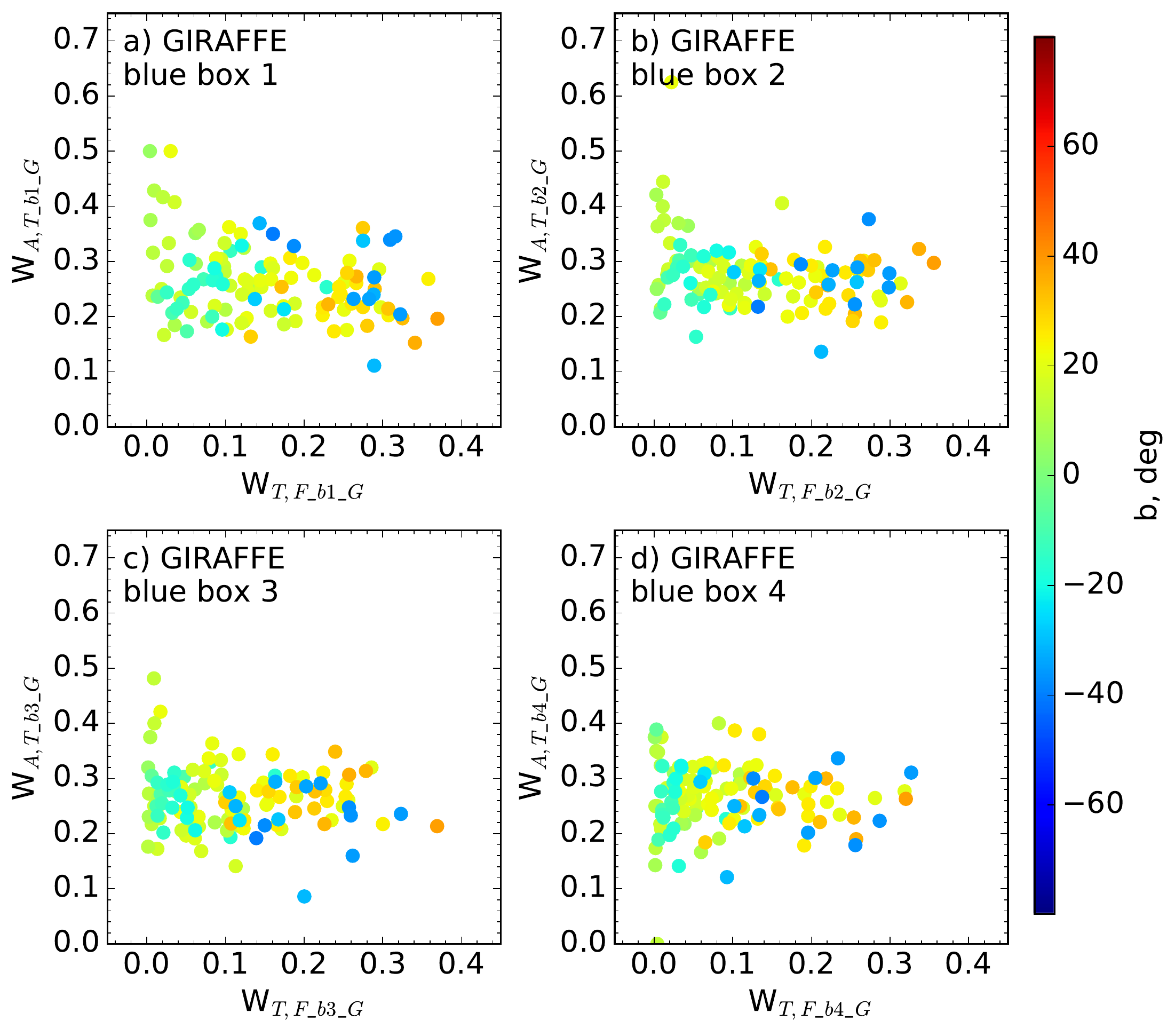}
\caption{Weights of each field for targeted stars versus stars in the field within a 1-degree FoV in diameter compared with weights for allocated versus targeted stars in blue box for GIRAFFE. (a)-(d) show weights in $J_{1-4}$ magnitude bins in fields within given FoV. The colour coding indicates Galactic latitude in degrees. 
}
\label{fig-weights3}
\end{figure}

\begin{figure}
\includegraphics[width=1\columnwidth]{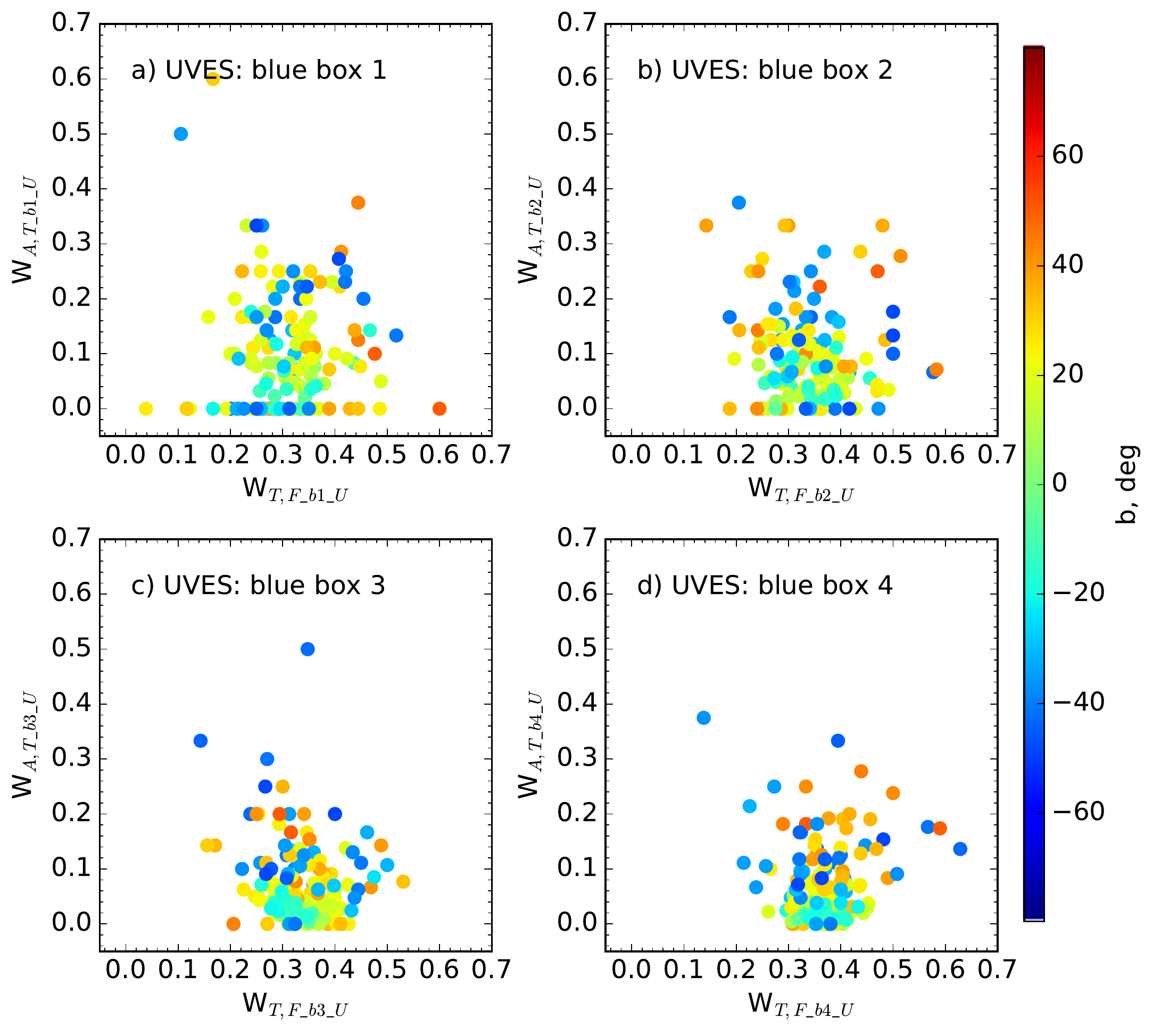}
\caption{Weights of each field for targeted stars versus stars in the field within a 1-degree FoV in diameter compared with weights for allocated versus targeted stars in blue box for UVES. (a)-(d) show weights in $J_{1-4}$ magnitude bins in fields within given FoV. The colour coding indicates Galactic latitude in degrees. 
}
\label{fig-weights4}
\end{figure}

\begin{figure*}
\resizebox{\hsize}{!}{
    \includegraphics[width=\columnwidth]{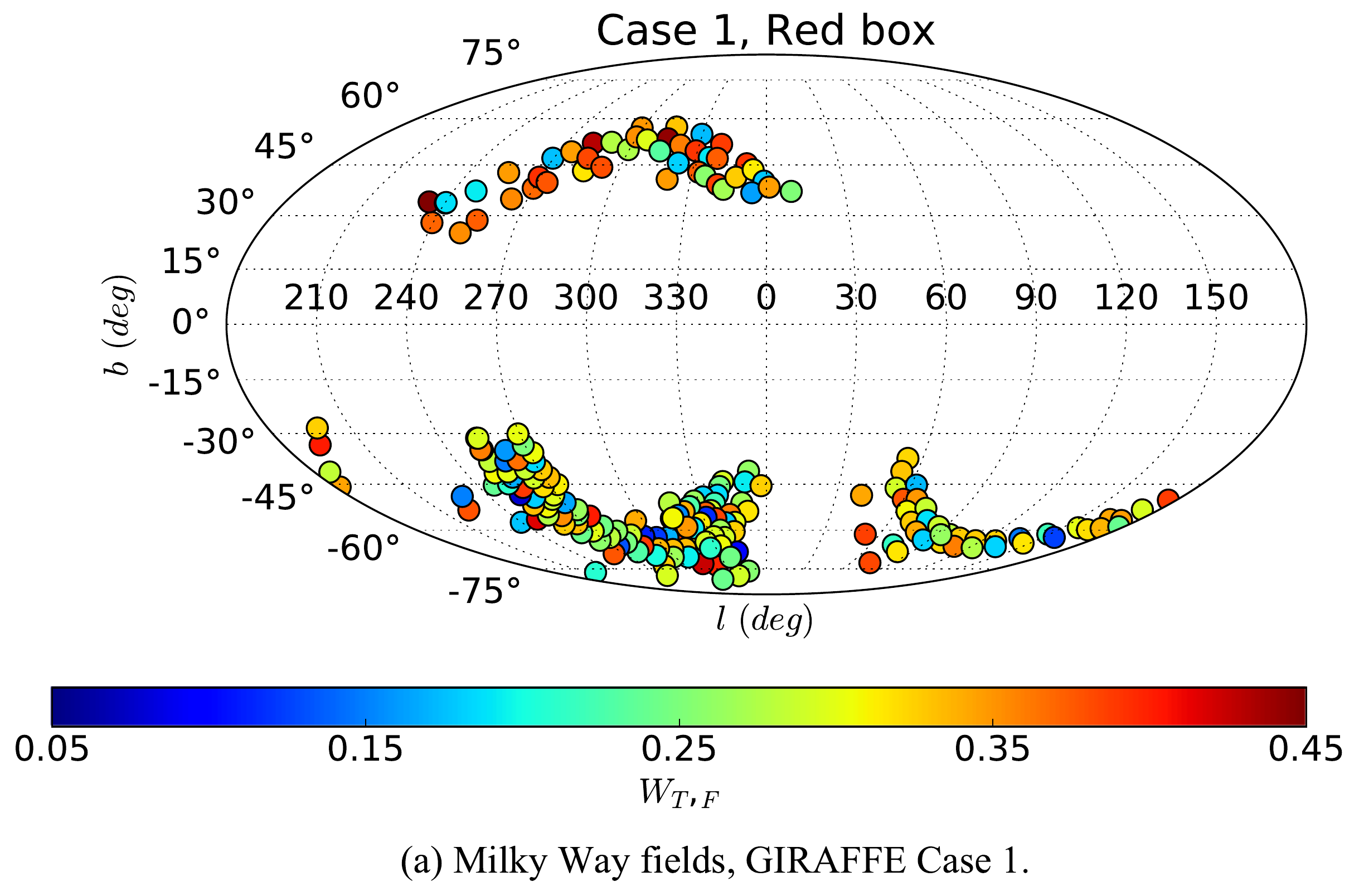}
      \includegraphics[width=\columnwidth]{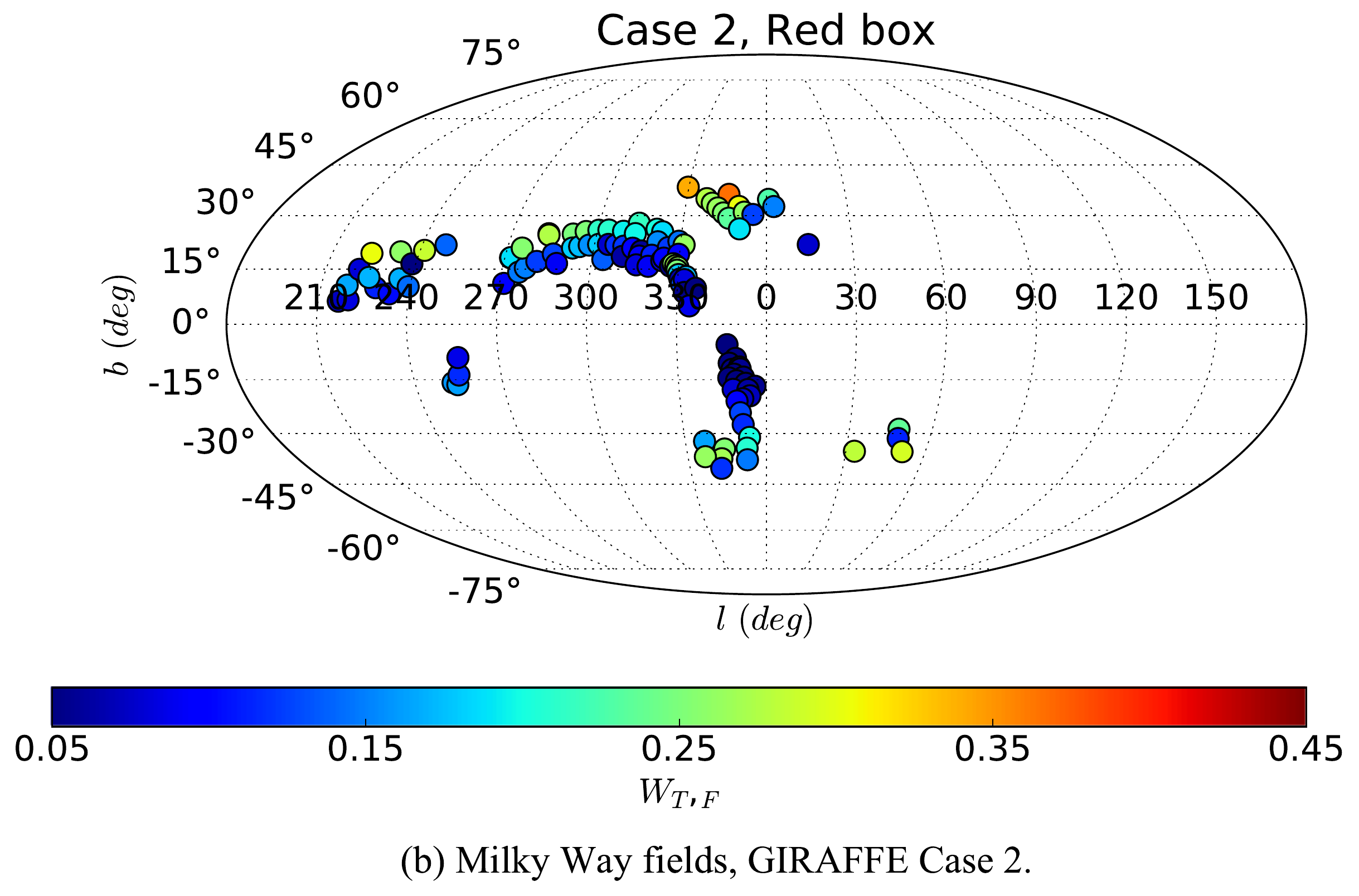}
}
\caption{The distribution, shown in Mollweide projection with the Galactic center in the middle, of the 
observed Milky Way fields across the sky. (a) and (b) show fields selected within the red box based on Case~1 and~2 respectively, and observed with GIRAFFE. The colour coding indicates the weight $W_{T,F}$ (Eq.~\ref{eq:WT}).}
\label{fig:W_red_box}
\end{figure*}


\section{Weights}\label{sec:Weights}

\begin{figure}
  \centering
  \includegraphics[width=1.\columnwidth]{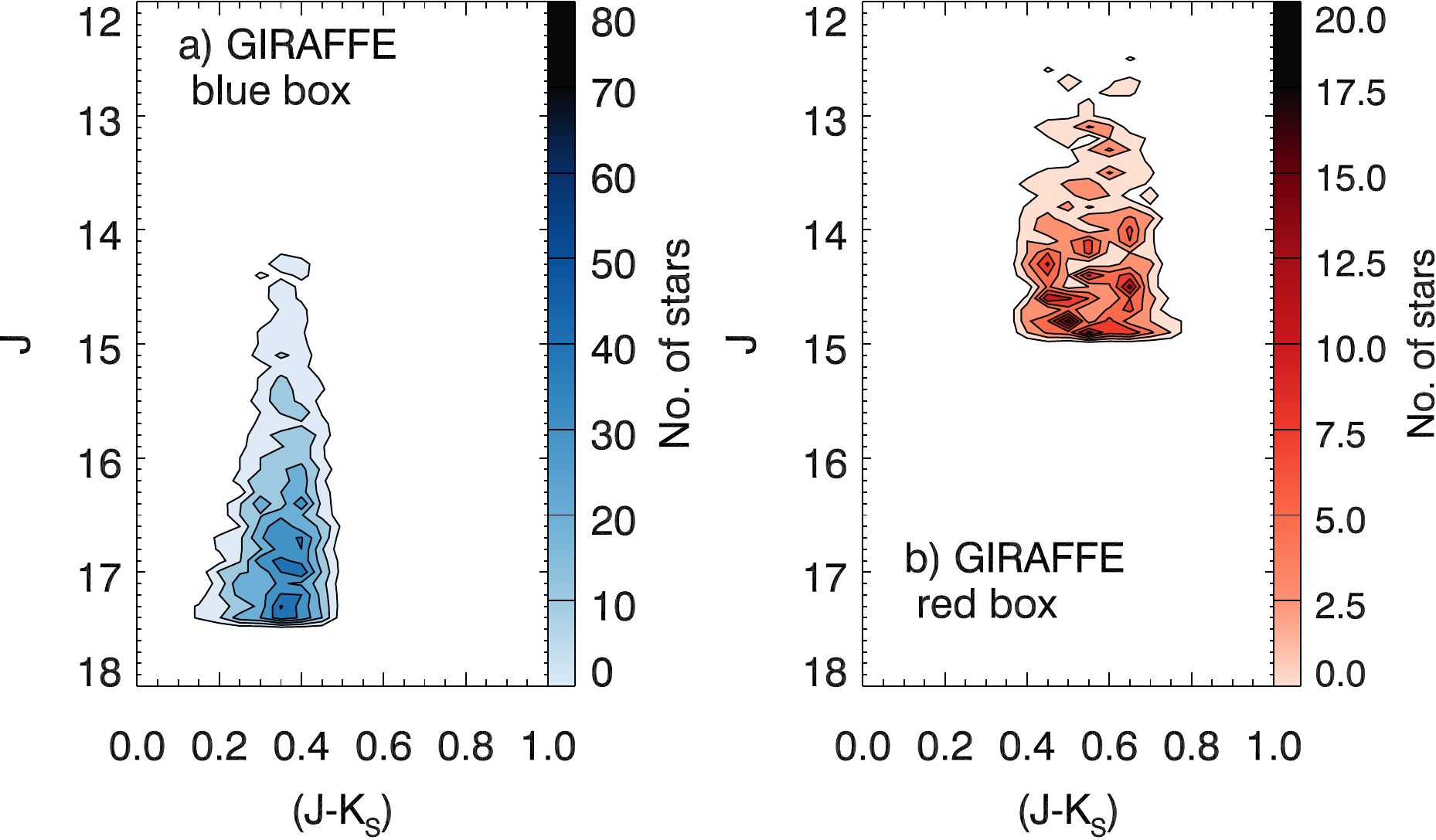}
  \caption{Distribution of the Milky Way field stars observed in iDR4, GIRAFFE that did not receive recommended stellar parameters in the same data release. 
  These are split into the blue (a) and red (b) boxes. }
  \label{fig:noparams}
\end{figure}

\subsection{Targeted and allocated weights}\label{sec:Weights1}

The selection function presented here consists of two steps.
The first step is where potential targets are selected for GIRAFFE and UVES. The second step, final target allocation, is generating the actual list for observation.

Here we present the weights per field calculated after the target selection and allocation. These weights can be used to better understand the {\it{Gaia}}-ESO Survey results and correct them for selection bias. 
 
The general  weight per field for the primary target selection is:
\begin{equation}
W_{T,F}=\frac{N_{T}}{N_{F}},
\label{eq:WT}
\end{equation}
where $N_{T}$ is the number of targeted objects in the field within 35$^\prime$ FoV in diameter. $N_{F}$ is the number of objects in the field within a 1-degree FoV in diameter (see Fig.~\ref{fig-FoV}). $W_{T,F}$ is the weight of targeted objects versus objects in the 1-degree FoV in diameter field.
To count $N_{F}$ for GIRAFFE targets we used the latest version of the VISTA VHS catalog (version~2015-04). 
We use the same VHS quality flags as in Table \ref{tab:flags} except for the flag~(iv) (VHS not on the bad {\it{CCD}}). 

The general  weight per field for the final target selection is:
\begin{equation}
W_{A,T}= \frac{N_{A}}{N_{T}},
\end{equation}
where $N_{A}$ is the number of allocated objects in the field within the FLAMES 25$^\prime$ FoV in diameter. $W_{A,T}$ is the weight of allocated objects versus targeted for a given Milky Way field.

Since the target selection function is complex, we calculated weights for all the CMD colour boxes separately (i.e. blue, red and extra) (Figs.~\ref{fig-weights1} and \ref{fig-weights2}). For Case~2 we calculated the blue box weights per $J_{1-4}$ magnitude bins in each field within the given FoV (Figs.~\ref{fig-weights3} and \ref{fig-weights4}). Hereafter, $J_{1-4}$$=$($J_{max}-J_{min}$)/4, where $J_{1}$ is the bright limit, and $J_{4}$ is the faint limit of the $J$ magnitude.   
All calculated CMD weights per field are listed in Table~\ref{table:online_T1}.

To illustrate the importance of accounting for the selection biases for individual Milky Way fields we show an example of the weight of the red box ($W_{T,F}$) distribution in the Milky Way fields (see Fig.~\ref{fig:W_red_box}).  
Case~1 Milky Way fields have  higher $W_{T,F}$ values than the Case~2 fields, but at the same time, the $W_{T,F}$ values are different for each individual field within the two cases.

In order to use Milky Way field stars for a specific science question, we must understand how the spectroscopic sample is drawn from the underlying population. 
As can be seen from Fig.~\ref{fig:W_red_box} the completeness  of the {\it{Gaia}}-ESO Survey varies substantially between fields. 
For each field we must therefore assess how representative the spectroscopic sample is of the underlying population. 
To correct for these types of biases the presented weights, $W_{T,F}$; $W_{A,T}$, should be used. 

\begin{figure*}
\resizebox{\hsize}{!}{
	\includegraphics[width=0.65\columnwidth]{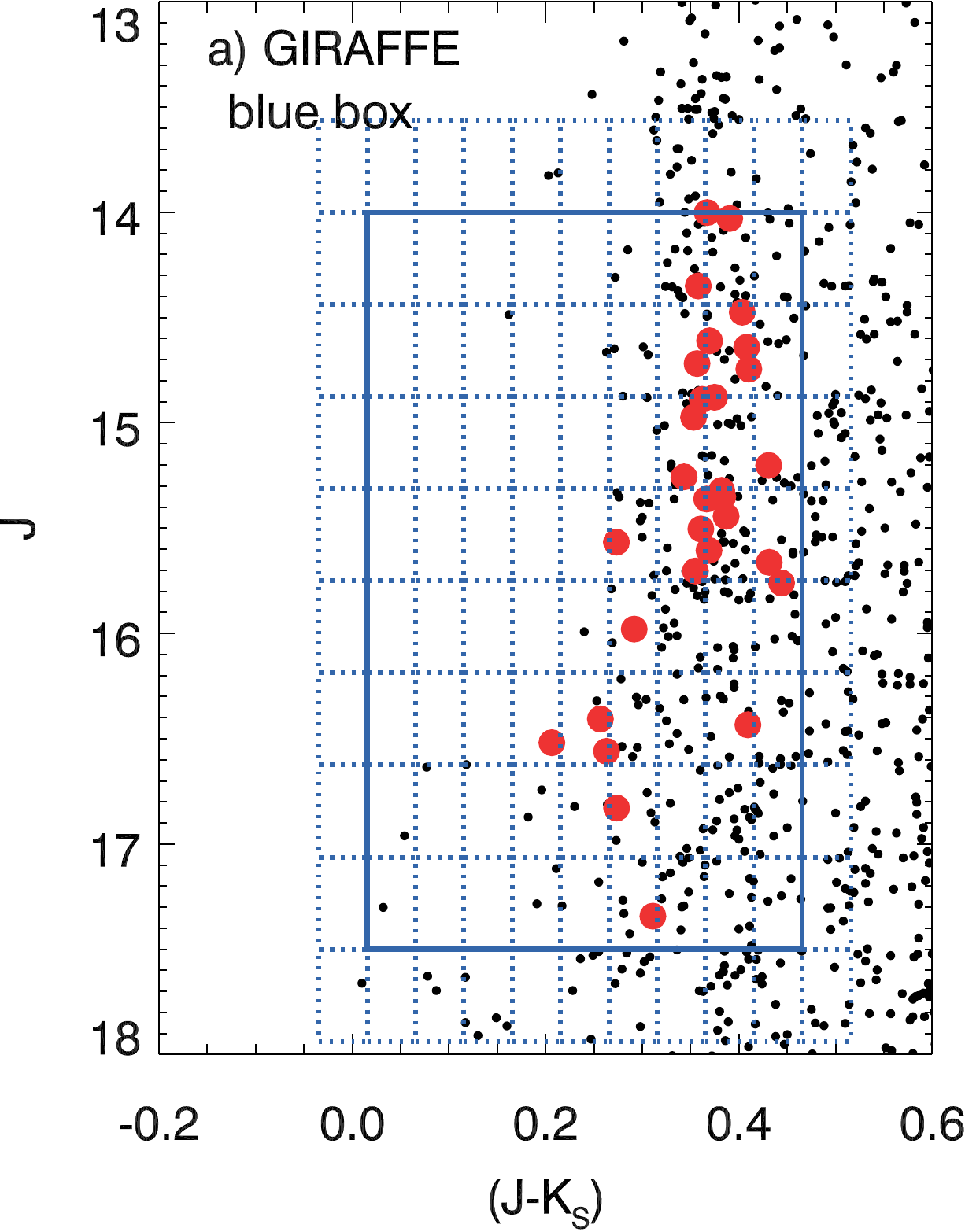}
	\includegraphics[width=0.65\columnwidth]{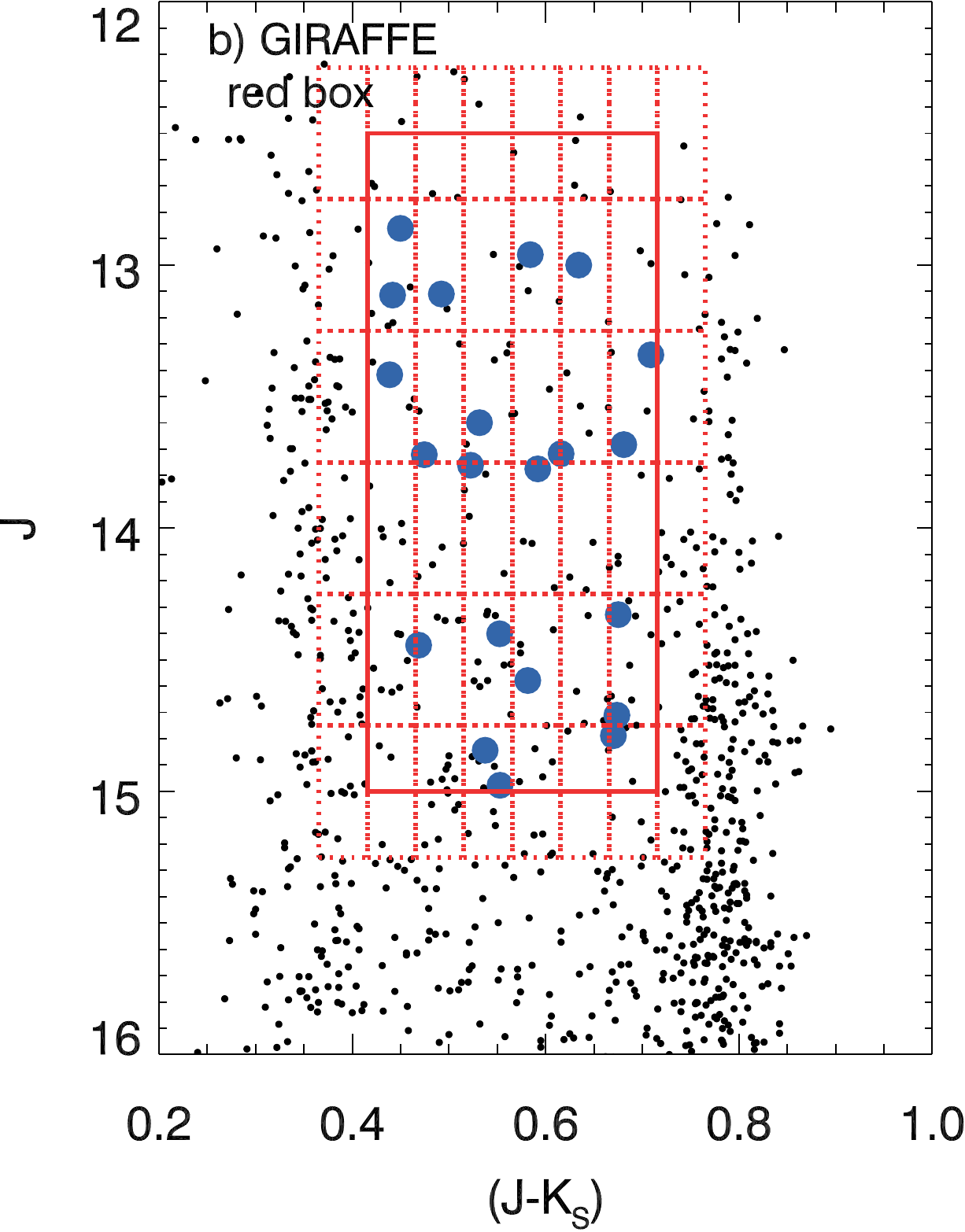}
	\includegraphics[width=0.65\columnwidth]{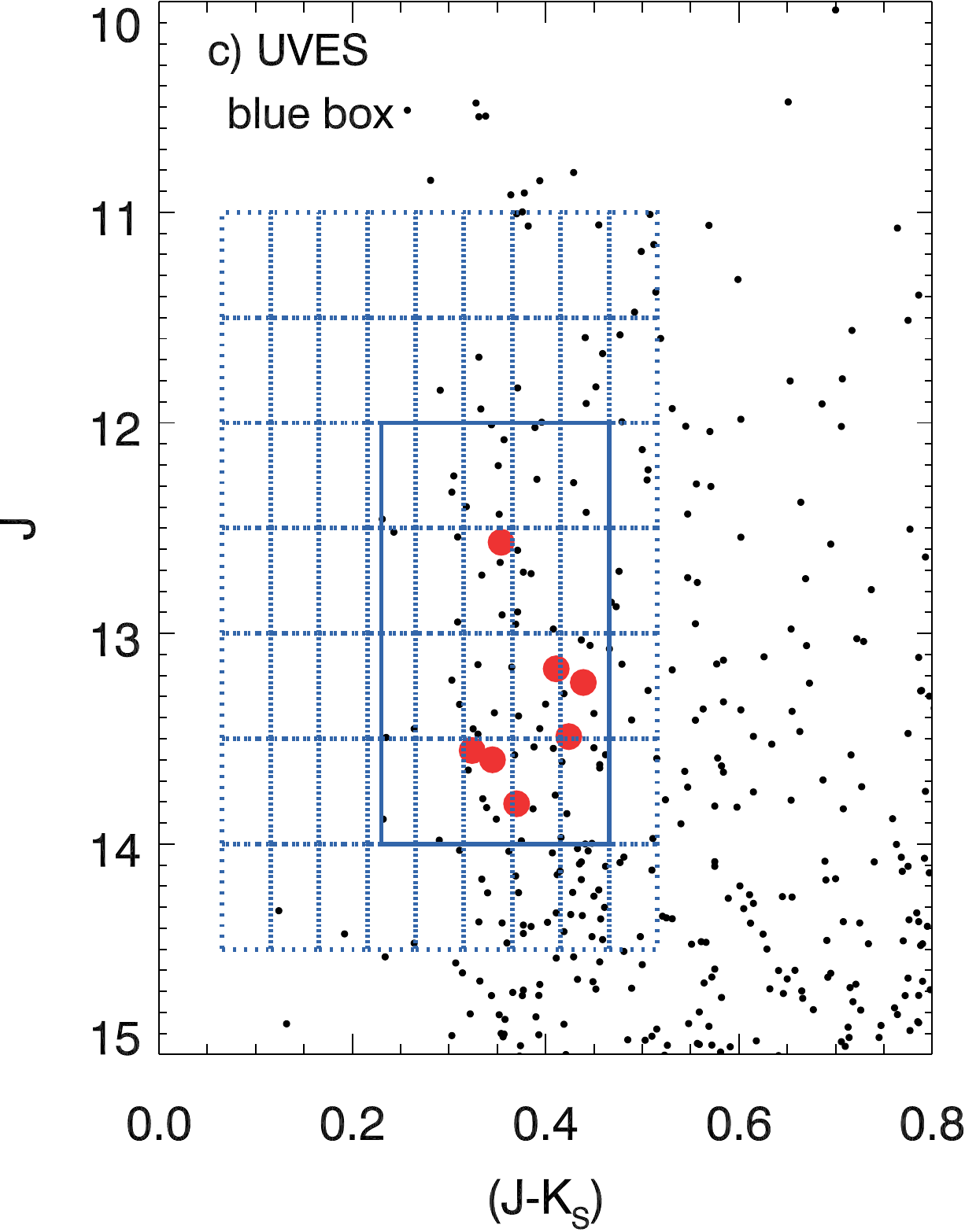}
	}
\caption{The colour-magnitude diagrams of an example of Milky Way field (GES\_MW\_023959-560000) for both the blue (a) and red~(b) boxes for GIRAFFE and (c) blue box for UVES. In (a) and (b) the background VISTA VHS and (c) 2MASS photometry (black points) and successfully observed targets (red and blue points) are shown. The red and blue solid lines outline the boxes used to select the targets, and the dashed lines show the grid of bins for weighting. }
\label{fig-idr4_grids}
\end{figure*}

\subsection{Using iDR4: stellar weights for the colour-magnitude diagram}\label{sec:weights_intro}

The selection function presented in this paper corrects for the discrepancy between the number of stars allocated to be observed and the number of stars originally available from the photometry for each field. This enables any comparison of fields to account for the varying population densities associated with different lines-of-sight.

In order to ensure a completely fair comparison of the data, however, a second correction is needed. Within each field, the density of stars available for observation varies considerably with respect to both colour and magnitude. Furthermore, not all observed stars end up with reasonable parameters; a significant proportion of observations fail to produce high enough quality spectra to enable robust stellar parameter determination. Naturally, the fainter targets are more likely to fail, due to the lower S/N of the spectra obtained. This is shown clearly in Fig.~\ref{fig:noparams}, where all stars observed in iDR4 that failed to result in stellar parameters are plotted on the colour-magnitude diagram. There are 3849 stars in the blue box without parameters (e.g. T$_{eff}$, log(g), [Fe/H]), and 856 in the red box.

To correct for these biases, each star that has values for the recommended stellar parameters in iDR4 needs a second weighting. This will ensure that results from the data release can be properly interpreted in terms of the actual populations of the Milky Way. 

\subsubsection{The CMD grids}
To calculate the weights for iDR4 stars, we divide the colour-magnitude diagram into a grid of bins. The bin size is sufficiently small to accurately reflect the local sampling around each observed star.
An example field is shown in Figs.~\ref{fig-idr4_grids}a and b, and it can be seen that the grid is larger than the original selection box. We are using the latest VISTA VHS photometry to calculate these weights, however many of the fields were observed some time ago, and the selection was completed with an older version of the VISTA VHS catalogs.
The magnitudes in the updated catalogs differ from the older catalogs by a very small amount, but these differences are enough to mean that some stars which previously fell inside the selection box now lie slightly outside. The larger grid allows us to include weights for those stars as well. 
 The red box is divided into bins of size 0.05 in ($J-K_{S}$), and 0.5 in $J$. The grid for the blue box was designed to overlap with the four magnitude boxes that were defined in Section \ref{subsec:case2} for Case~2 fields. Therefore the bins are 0.4375 long in $J$ (half the size of the magnitude boxes $J_{1-4}$), and again 0.05 wide in ($J-K_{S}$).

A similar set-up is used for the UVES blue box (see Fig.~\ref{fig-idr4_grids}c) as in the GIRAFFE blue box, with different sized magnitude bins again to match the boxes in Case~2 fields; the bins are 0.5 long in $J$, and remain 0.05 wide in ($J-K_{S}$). In order to cope with those fields which had a bright limit of $J_{2MASS}=11$ rather than $J_{2MASS}=12$, the grid has been extended, as shown in Fig.~\ref{fig-idr4_grids}c.

 \begin{table*}
\centering
  \caption{CMD weights of successfully observed stars with GIRAFFE in iDR4. The full table is available online.}
  \begin{tabular}{cccccccccccccccccccccccccccccccc}
    \hline
     CNAME& GES\_FLD & RA[deg] & Dec[deg] & W\_O,F\_G  & W\_Total\_G & Box\\
    \hline
  00000301-5455591 & GES\_MW\_000024-550000 & 0.0125 & -54.9331 & 0.1333333 & 0.0173717 & Blue\\
  00000377-5506384 & GES\_MW\_000024-550000 & 0.0157 & -55.1107 & 0.2000000 & 0.0260576 & Blue\\
  00000395-5458308 & GES\_MW\_000024-550000 & 0.0165 & -54.9752 & 1.0000000 & 0.1302880 & Blue\\
  00000533-5459505 & GES\_MW\_000024-550000 & 0.0222 & -54.9974 & 0.2500000 & 0.0325720 & Blue\\
  00000648-5451013 & GES\_MW\_000024-550000 & 0.0270 & -54.8504 & 0.1500000 & 0.0195432 & Blue\\
    ... & ... & ... & ... & ... & ... & ...\\
    \hline
  \end{tabular}
    \label{table:online_T2}
 \end{table*}
 
 \begin{table*}
\centering
  \caption{CMD weights of successfully observed stars with UVES in iDR4. The full table is available online.}
  \begin{tabular}{ccccccccccccccccccccccccccccccc}
    \hline
     CNAME& GES\_FLD & RA[deg] & Dec[deg] & W\_O,F\_U  &  W\_Total\_U \\
    \hline
  00000009-5455467 & GES\_MW\_000024-550000 & 0.0003 & -54.9296 & 0.20000 & 0.01692 \\
  00001749-5449565 & GES\_MW\_000024-550000 & 0.0728 & -54.8323 & 0.16667 & 0.01410 \\
  00012216-5458205 & GES\_MW\_000024-550000 & 0.3423 & -54.9723 & 0.14286 & 0.01208 \\
  00035430-0058050 & GES\_MW\_000400-010000 & 0.9762 & -0.9680   & 0.25000 & 0.00912 \\
  00035518-0047502 & GES\_MW\_000400-010000 & 0.9799 & -0.7972   & 0.33333 & 0.01217 \\
    ... & ... & ... & ...  & ...  & ... \\
    \hline
  \end{tabular}
    \label{table:online_T3}
 \end{table*}

\begin{figure*}
\resizebox{\hsize}{!}{
	\includegraphics[width=\columnwidth]{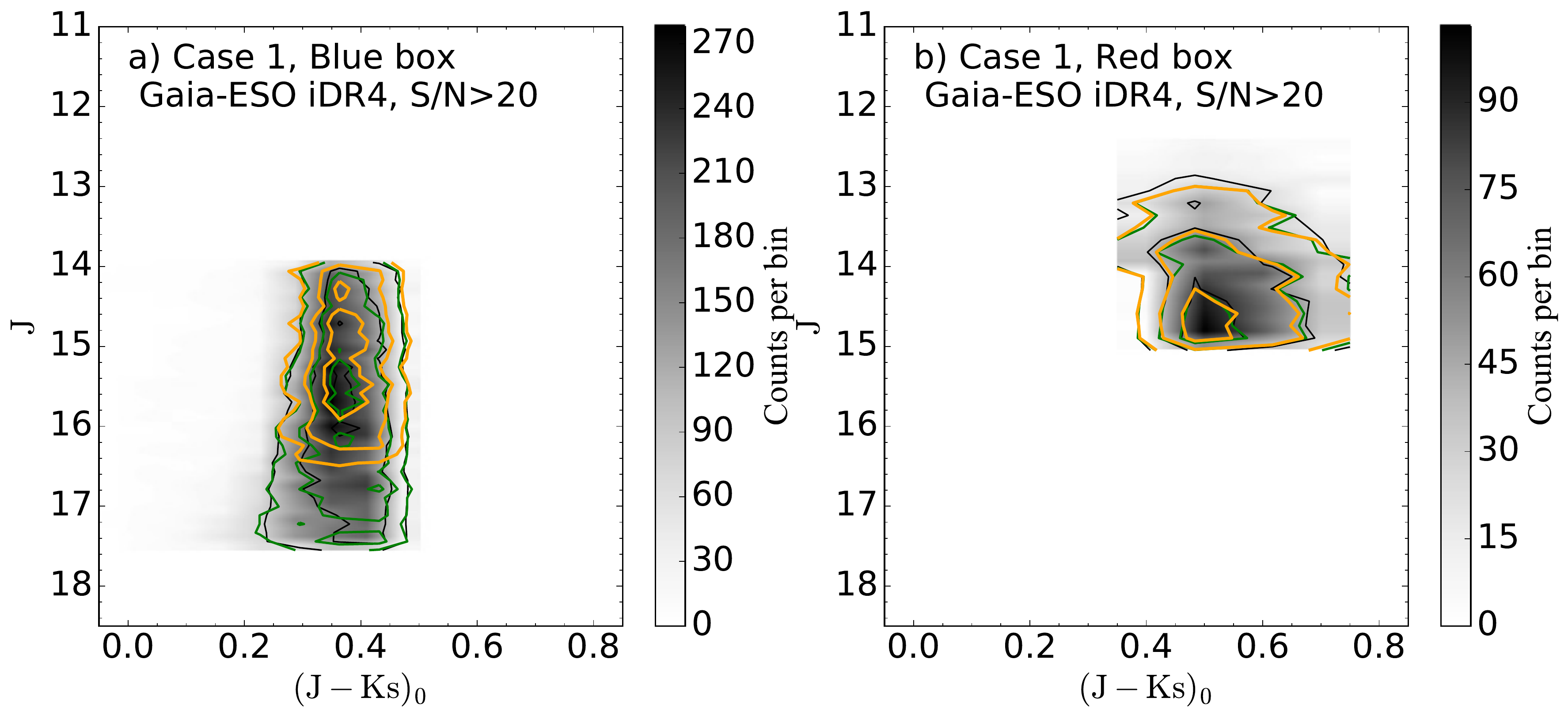}
	\includegraphics[width=\columnwidth]{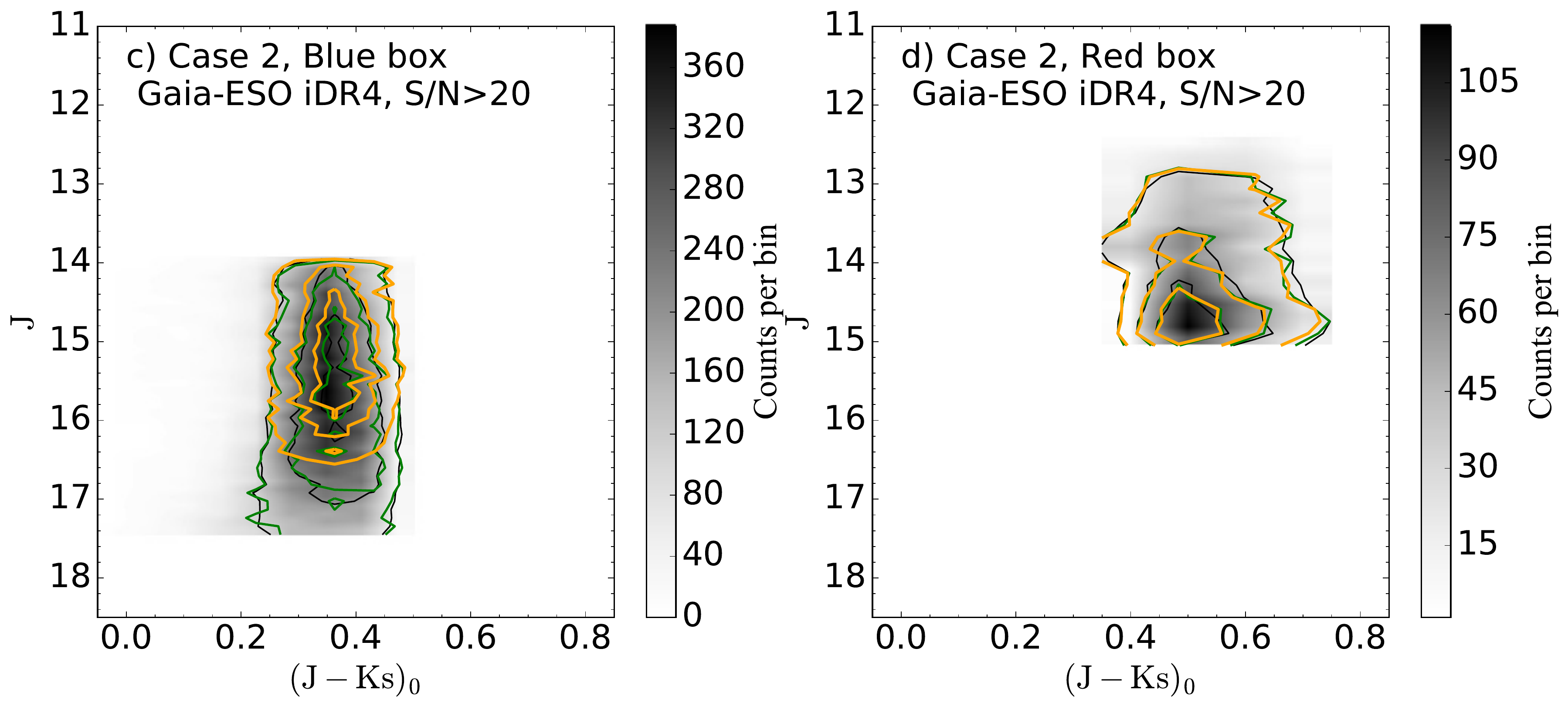}
	}
\caption{Distribution of the photometric sample of the {\it{Gaia}}-ESO Survey Milky Way fields in blue (a,c) and red (b,d) boxes (linear density grey scale, black contours).  The observed spectroscopic sample of the {\it{Gaia}}-ESO Survey iDR4 is shown as green contours; yellow contours show the spectroscopic sample with determined effective temperature and the signal-to-noise ratio cut of S/N $>$ 20. The black, green and yellow contours contain 68$\%$, 95$\%$, and 99$\%$ of the distribution of Case~1 and Case~2  Milky Way field stars. On the x-axis we show the de-reddened (J-K$_{s}$)$_0$ colour, whereas on the y-axis we show the observed $J$ magnitude.} 
\label{fig-idr4_r_box_grey}
\end{figure*}

\begin{figure*}
\resizebox{.85\hsize}{!}{
    \includegraphics[width=0.4\columnwidth]{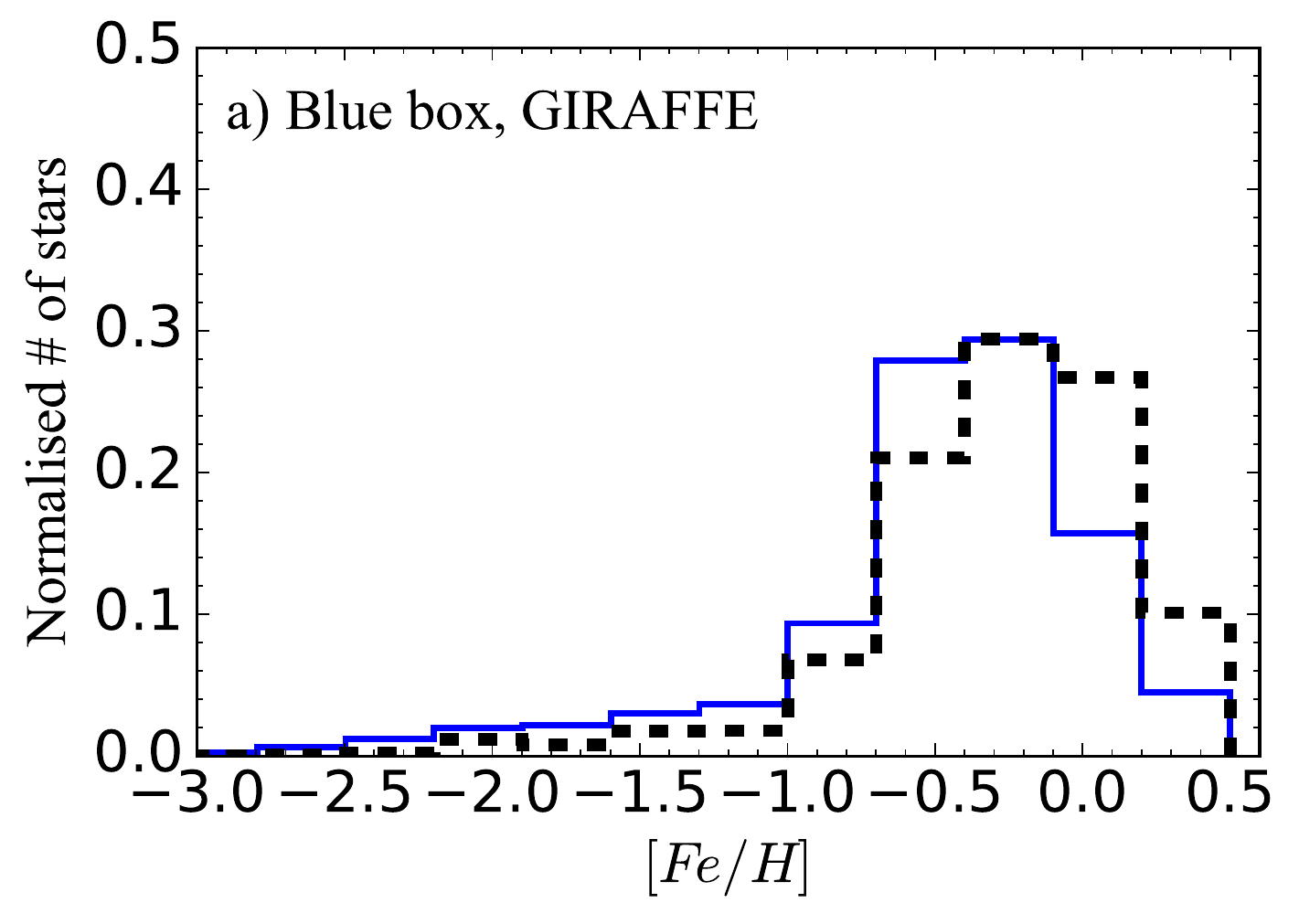}
    \includegraphics[width=0.4\columnwidth]{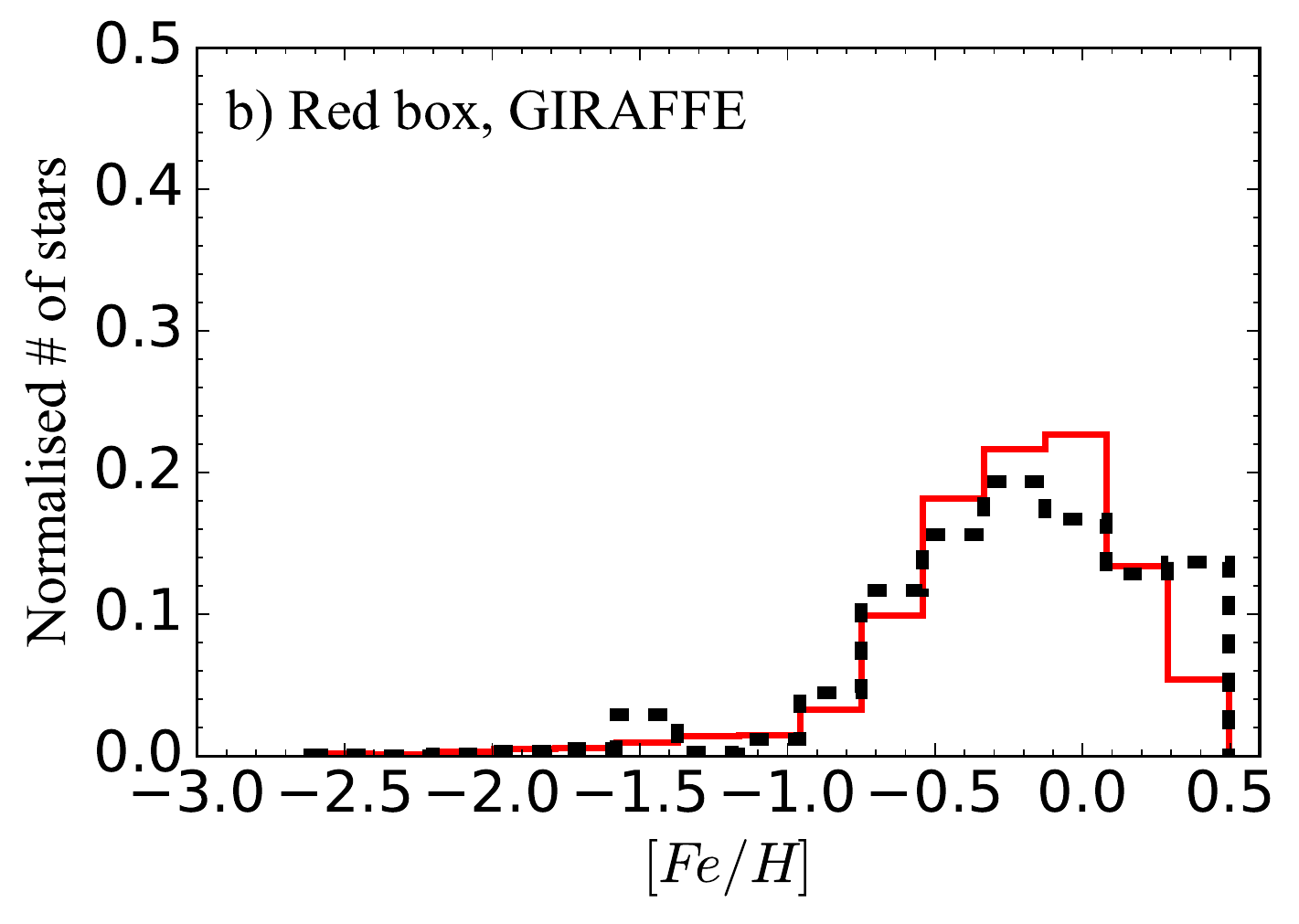}
}
\resizebox{.85\hsize}{!}{
    \includegraphics[width=0.4\columnwidth]{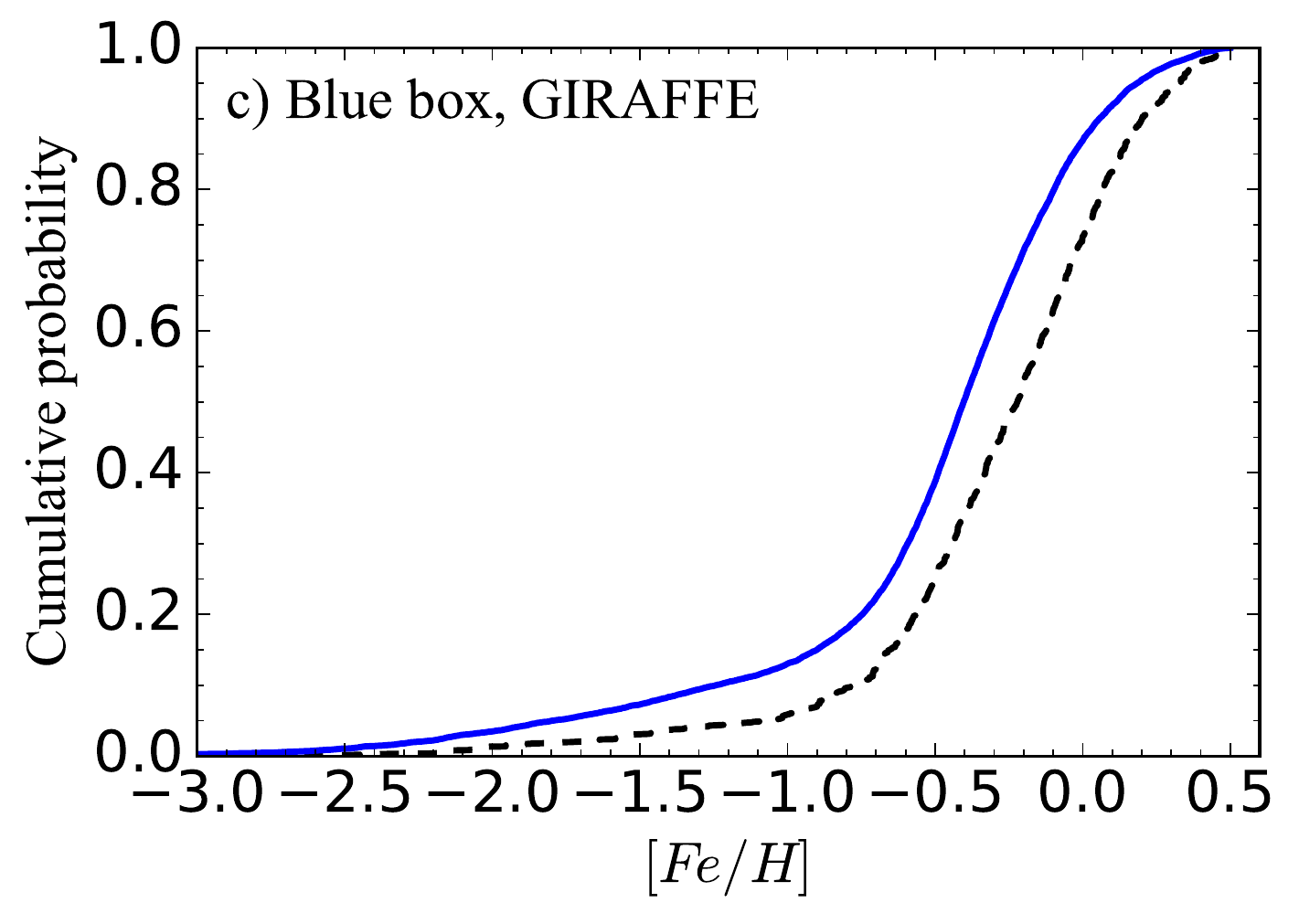}
    \includegraphics[width=0.4\columnwidth]{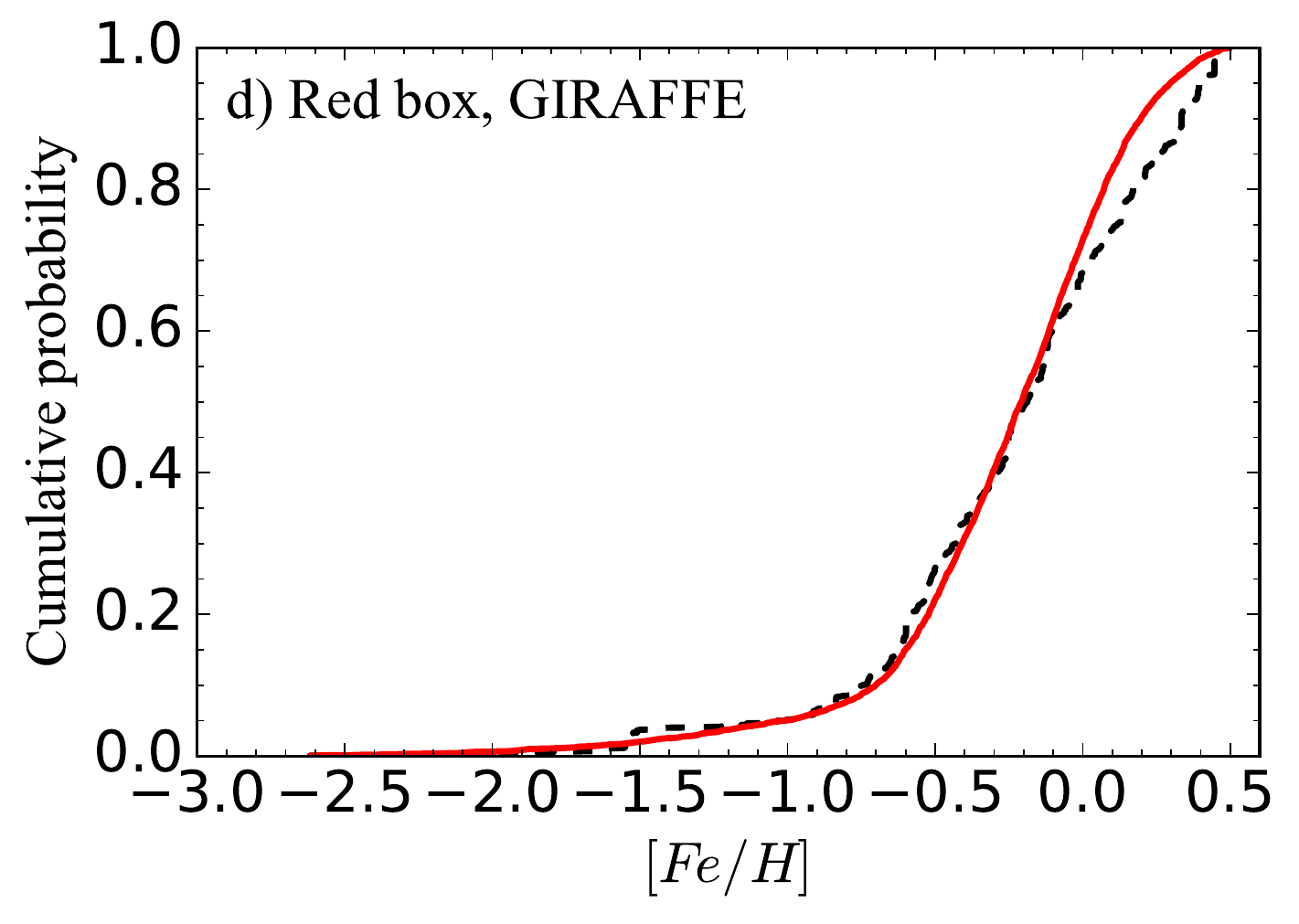}
}
\caption{The metallicity distribution of all Milky Way field stars observed with GIRAFFE in the {\it{Gaia}}-ESO Survey iDR4. (a, c) and (b, d) show metallicity distribution of stars within blue and red boxes respectively. 
The blue and red lines show the observed metallicity; the black dashed lines show weighted metallicity distributions. Only stars with $[Fe/H] < 0.50 $ are included. }
\label{fig:weights_2}
\end{figure*}

\subsubsection{ Weights of successfully observed targets}\label{sec:Weights2}

The weight of a successfully observed target in each CMD bin is calculated as follows:
\begin{equation}
W_{O_{bin},F_{bin}}=\frac{N_{O_{bin}}}{N_{F_{bin}}},
\end{equation}
where $N_{O_{bin}}$ is the number of successfully observed targets in that bin, and $N_{F_{bin}}$ is the number of objects in the VISTA VHS or 2MASS photometry for that bin. It is important to note that $N_{O_{bin}}$ is not the same as $N_{A_{bin}}$, that is, the number of allocated objects in the bin, because $N_{O_{bin}}$ only counts those objects that were successfully observed and have parameters in iDR4. 
The weights $W_{O_{bin},F_{bin}}$ of successfully observed stars with GIRAFFE and UVES are listed in Tables~\ref{table:online_T2} and~\ref{table:online_T3} where ``CNAME'' is a {\it{Gaia}}-ESO surveys specific stellar ID.
The weights have not been calculated for those stars which fell into the extra box in Case~1 fields and not for SDSS and SkyMapper targets.
We decided not to compare the stars in the extra box with stars from other fields with  Case~1 selection, because targets within the extra box were selected with right-edge extensions varying between the fields (see Fig.~\ref{fig_delta_g_u}).

\section{ A first look at iDR4 successfully analysed data}\label{sec:idr4}

As we mentioned before the selection function consists of several steps. In addition to selection and allocation of targets we also need to know the
completeness of the successfully analysed {\it{Gaia}}-ESO Survey Milky Way field sample, especially to understand the bias introduced by the signal-to-noise ratio variation in the observed FLAMES spectra.  Therefore we looked at the {\it{Gaia}}-ESO Survey iDR4 data. The internal DR4 is a full release. All observations from the beginning of the survey until July 2014 are included.

Not surprisingly, the signal-to-noise ratio varies with $J$~magnitude. 
We made a potential quality cut on S/N ratio of the spectra. Inspecting the spectroscopic results (e.g.~$T_{eff}$, [Fe/H]) we chose to cut the spectroscopic sample at a median S/N~>~20.
This cut might be different if one wants to consider other spectroscopic results e.g. alpha abundance. 
Figure~\ref{fig-idr4_r_box_grey} shows the relation between the potential photometric sample (black contours), the spectroscopic sample (green contours), and after the quality cut on signal-to-noise ratio (yellow contours) for the Milky Way field GIRAFFE sample. 
In our case, while the sampling in colour is close to unbiased, the sampling in $J$ is strongly biased against faint targets because of the signal-to-noise cut (S/N $>$ 20).
Similar trends are seen in the SEGUE G-dwarf sample analysed by \citet{bovy12}.
Introducing this quality cut we lose about one magnitude in depth for targets within the blue box, while the sampling in the red box is not that affected (see Fig.~\ref{fig-idr4_r_box_grey}). 

Here we provide a simple example of how one can use the presented weights.
We select Milky Way field stars observed with GIRAFFE from iDR4 data within blue and red boxes and with [Fe/H]~<~0.50 (i.e. the more metal-rich targets are very uncommon). Here we chose to look at GIRAFFE targets only and show an application of the presented weights but in principal the calculated weights can be applied to UVES targets as well.  These weights can be applied on top of the selection function weights described in Section \ref{sec:Weights1}, by multiplying them together as follows: 
\begin{equation}
W_{Total}={W_{A,T}}\times{W_{T,F}}\times{W_{O_{bin},F_{bin}}},
 \label{eq:w_total}
\end{equation}
where $W_{Total}$ is the total weight. The total weight $W_{Total}$ of a successfully observed star with GIRAFFE and UVES are listed in Tables~\ref{table:online_T2} and~\ref{table:online_T3}.

Stars observed with GIRAFFE that are in the region of the CMD where blue and red boxes overlap will have two $W_{Total}$ weights and will be indicated in Table~\ref{table:online_T2} as Blue/Red (calculated as blue box star) or Red/Blue (calculated as red box star) targets. Our recommendation is to use the blue box weight for those stars (607 stars in iDR4) when combining blue and red box data.

We can therefore, when trying to accurately sample the Milky Way disc, characterise the importance of an observed star in iDR4 by giving it a weight.
In order to make a correction for each successfully observed star in our sample we give the total weight $W_{Total}$ as 1/$W_{Total}$.
Here we effectively tell how frequent a successfully observed stars is with a given $J$ and ($J-K_{s}$) within a given Milky Way field in a given FoV.
In this example we chose to remove stars where 1/$W_{Total}$~>~100 000, because in this case the weight is too large to be meaningful, and the star
is not representative of the sample. Essentially, this cut removes 1.2\% of blue box targets and 0.2\% of red box targets. 

In Fig.~\ref{fig:weights_2} we show an application of the weights presented in this paper. We see that the metallicity distribution of stars observed in the {\it{Gaia}}-ESO Survey iDR4 with GIRAFFE in the blue and red boxes (blue and red line) are different from the weighted distributions (black dashed line) when all Milky Way fields are analysed together. 
We find that for the red box the observed distribution is the same as the underlying distribution, but this is not the case for the blue box, as seen by comparing their respective cumulative distributions, corrected vs. uncorrected (see  Fig.~\ref{fig:weights_2}c and d).
The total weight,  $W_{Total}$, normalises between the different lines of sight; while each observed Milky Way field has almost the same number of spectra, but not the same number of successfully analysed stars and different number of photometric objects varying per line-of-sight due to the nature of the Galaxy. This is a simple example to highlight the necessity of including such information in studies of the stellar populations in the Milky Way. There are other ways how one can apply presented weights (i.e. taking into account the [Fe/H] errors, looking at different lines-of-site or combining blue and red box targets together). 

The presented field CMD weights ($W_{A,T}$, $W_{T,F}$) and the weights of successfully observed targets ($W_{O_{bin},F_{bin}}$) can be used differently than in the previous example. For example, looking at the radial and/or vertical metallicity distribution 
the weights can be used to limit the data to only those stars that most represent the underlying population.  In this case, we do not want stars with very small $W_{Total}$ to contribute to the analysis, since they do not provide enough information about the actual underlying population. A simple way of studying the Milky Way's radial metallicity gradient is to bin the data in Galactocentric radial distance R and compute the mean metallicity in each bin as a running average. However, instead of computing a straight mean, we can perform a weighted average, in which each star is weighted by $W_{Total}$.  This will then bias the mean metallicity towards those stars with the highest $W_{Total}$. The results can then be compared with those for the standard mean to understand possible biases in the data.

\section{Conclusions and future prospects}\label{sec:Conclusions}

We have discussed the details of the selection function for the Milky Way field stars observed in the {\it{Gaia}}-ESO Survey. The weights presented here are based on targets selected from the beginning of the survey up to the end of June 2015.
To characterise the major components of the Galaxy, and to understand these components in the context of the Milky Way's formation and evolution history, the survey selection function is designed to target stars as homogeneously as possible throughout  the Milky Way.
The target selection is based on stellar magnitudes and colours, using photometry from the VISTA Hemisphere Survey \citep{mcmahon13} and 2MASS \citep{skrutskie06}. Also we present the basic and actual target selection schemes for the Milky Way field stars observed with FLAMES/GIRAFFE and FLAMES/UVES.

The actual target selection scheme is divided into two cases. In Case~1 the target selection algorithm, in addition to the two main selection CMD boxes (i.e. blue, red), extends the colour limits to select second priority targets (i.e. for those Milky Way fields where the density of stars is not enough to fill the FLAMES fibres).
Case 2 is used to select targets near the Galactic plane. In this case the target selection algorithm is configured to select the same number of targets per magnitude bin (i.e. not to have a bias towards very faint stars). 

From the beginning of the survey on December 31, 2011 until June 2015,  a total of 330 Milky Way fields have been targeted and allocated on FLAMES. 202~Milky Way fields were selected using Case~1 and 118~using Case~2, which were then allocated on FLAMES/GIRAFFE.  
For UVES,~164 and 166~Milky Way fields were used in Case~1 and Case~2, respectively. In addition, a sample of Milky Way fields were selected to target rare but astrophysically important stellar populations (e.g. metal-poor stars, K~giants), where Milky Way field targets were selected using the SDSS photometry \citep{ahn12} and SkyMapper photometry \citep{keller07}, for allocation on FLAMES/GIRAFFE and FLAMES/UVES. 

The {\it{Gaia}}-ESO survey selection function depends not only on potentially selected targets but also on allocated targets. It is crucial to know the number of stars that were not allocated to any spectroscopic fibre, i.e., FLAMES/GIRAFFE fibres, in order to afterwards correct for any incompleteness effects  on the survey.
Finally, we presented the weights calculated after the target selection and allocation. These weights can be used to better understand the {\it{Gaia}}-ESO Survey results and correct selection biases for the proper interpretation of the data in terms of our understanding of the Milky Way as a galaxy. 

We are continuing our work on weights application for the {\it{Gaia}}-ESO Survey Milky Way field GIRAFFE data and the results will be presented in a forthcoming paper. Here we presented weights per field for targeted and allocated Milky Way stars observed up to end of June, 2015 and weights of successfully observed iDR4 stars (15\,154 for GIRAFFE and 1\,367 for UVES) and we plan to continue our work on the target selection function for Milky Way stars when the next data set is available.

{\small{\section*{Acknowledgements}
ES, LH, SF, GRR and TB acknowledge support from the project grant ``The New Milky Way'' from the Knut and Alice Wallenberg Foundation. 
RS acknowledges support from NCN/Poland through grant 2014/15/B/ST9/03981. AJK acknowledges support by the Swedish National Space Board. We thank the anonymous referee for her/his constructive comments, which have improved the clarity of the paper.
Based on data products from observations made with ESO Telescopes at the La Silla Paranal Observatory under programme ID 188.B-3002. These data products have been processed by the Cambridge Astronomy Survey Unit (CASU) at the Institute of Astronomy, University of Cambridge, and by the FLAMES/UVES reduction team at INAF/Osservatorio Astrofisico di Arcetri. These data have been obtained from the {\it{Gaia}}-ESO Survey Data Archive, prepared and hosted by the Wide Field Astronomy Unit, Institute for Astronomy, University of Edinburgh, which is funded by the UK Science and Technology Facilities Council.
This work was partly supported by the European Union FP7 programme through ERC grant number 320360 and by the Leverhulme Trust through grant RPG-2012-541. We acknowledge the support from INAF and Ministero dell' Istruzione, dell' Universit\`a' e della Ricerca (MIUR) in the form of the grant ``Premiale VLT 2012''. The results presented here benefit from discussions held during the {\it{Gaia}}-ESO workshops and conferences supported by the ESF (European Science Foundation) through the GREAT Research Network Programme.
Based on observations obtained as part of the VISTA Hemisphere Survey, ESO Program, 179.A-2010 (PI: McMahon).
Funding for SDSS-III has been provided by the Alfred P. Sloan Foundation, the Participating Institutions, the National Science Foundation, and the U.S. Department of Energy Office of Science. The SDSS-III web site is http://www.sdss3.org/.}}




\bibliographystyle{mnras}
\bibliography{Stonkute_v2MNRAS} 


\appendix

\section{Calculated CMD weights}

For each Milky Way field observed from the beginning of the {\it{Gaia}}-ESO Survey until June 2015 we calculated CMD weights per field $W_{T,F}$ and $W_{A,T}$ to correct for potential biases. Those Milky Way fields and associated field weights are in Table~\ref{table:online_T1} and the meanings of acronyms used are in Table~\ref{table:online_log}. Full version of Table~\ref{table:online_T1} is available online.
 
\begin{table*}
\centering
\small
\caption{The meanings of acronyms used in Table~\ref{table:online_T1}.}
\begin{tabular}{lllc}
\hline
Col$\_$No & Acronym & Meaning \\
\hline
\hline
(1)&GES$\_$FLD   & GES Milky Way field name from CASU\\
(2)&RA  & RA [h:m:s] of GES Milky Way field center \\
(3)&Dec & Dec [d:m:s] of GES Milky Way field center \\
(4)&E(B$-$V)		& The Galactic dust extinction median value measured from the Schlegel, Finkbeiner \& Davis (1998) maps\\
(5)&$\Delta_{G}$	& The right edge limit of second priority targets for GIRAFFE\\
(6)&$\Delta_{U}$	&The right edge limit of second priority targets for UVES\\
(7)&W$\_$T,F$\_$b$\_$G	&	The weight of blue box for GIRAFFE, where targeted star counts are versus star counts in 1$^\circ$~FoV\\		
(8)&W$\_$T,F$\_$b1$\_$G	&	The weight of blue box with $J1$ for GIRAFFE, where targeted star counts are versus star counts in 1$^\circ$~FoV\\
(9)&W$\_$T,F$\_$b2$\_$G	&	The weight of blue box  with $J2$ for GIRAFFE, where targeted star counts are versus star counts in 1$^\circ$~FoV\\
(10)&W$\_$T,F$\_$b3$\_$G	&	The weight of blue box  with $J3$ for GIRAFFE, where targeted star counts are versus star counts in 1$^\circ$~FoV\\
(11)&W$\_$T,F$\_$b4$\_$G	&   	The weight of blue box  with $J4$ for GIRAFFE, where targeted star counts are versus star counts in 1$^\circ$~FoV\\    
(12)&W$\_$T,F$\_$r$\_$G	&   	The weight of red box  for GIRAFFE, where targeted star counts are versus star counts in 1$^\circ$~FoV\\
(13)&W$\_$T,F$\_$e$\_$G	&   	The weight of extra box for GIRAFFE, where targeted star counts are versus star counts in 1$^\circ$~FoV\\
(14)&W$\_$A,T$\_$b$\_$G  	&	The weight of blue box for GIRAFFE, where allocated star counts are versus targeted star counts\\
(15)&W$\_$A,T$\_$b1$\_$G 	&	The weight of blue box with $J1$ for GIRAFFE, where allocated star counts are versus targeted star counts\\         
(16)&W$\_$A,T$\_$b2$\_$G 	&	The weight of blue box with $J2$ for GIRAFFE, where allocated star counts are versus targeted star counts\\        
(17)&W$\_$A,T$\_$b3$\_$G   	&	The weight of blue box with $J3$ for GIRAFFE, where allocated star counts are versus targeted star counts\\       
(18)&W$\_$A,T$\_$b4$\_$G   	&	The weight of blue box with $J4$ for GIRAFFE, where allocated star counts are versus targeted star counts\\       
(19)&W$\_$A,T$\_$r$\_$G   	&	The weight of red box for GIRAFFE, where allocated star counts are versus targeted star counts\\
(20)&W$\_$A,T$\_$e$\_$G	&	The weight of extra box for GIRAFFE, where allocated star counts are versus targeted star counts\\
(21)&W$\_$T,F$\_$b$\_$U	&		The weight of blue box for UVES, where targeted star counts are versus star counts in 1$^\circ$~FoV\\
(22)&W$\_$T,F$\_$b1$\_$U	&	The weight of blue box with $J1$ for UVES, where targeted star counts are versus star counts in 1$^\circ$~FoV\\
(23)&W$\_$T,F$\_$b2$\_$U	&	The weight of blue box with $J2$ for UVES, where targeted star counts are versus star counts in 1$^\circ$~FoV\\
(24)&W$\_$T,F$\_$b3$\_$U	&	The weight of blue box with $J3$ for UVES, where targeted star counts are versus star counts in 1$^\circ$~FoV\\
(25)&W$\_$T,F$\_$b4$\_$U     &	The weight of blue box with $J4$ for UVES, where targeted star counts are versus star counts in 1$^\circ$~FoV\\       
(26)&W$\_$T,F$\_$e$\_$U   	&	The weight of extra box for UVES, where targeted star counts are versus star counts in 1$^\circ$~FoV\\
(27)&W$\_$A,T$\_$b$\_$U  	&	The weight of blue box for UVES, where allocated star counts are versus targeted star counts\\
(28)&W$\_$A,T$\_$b1$\_$U     &	The weight of blue box with $J1$ for UVES, where allocated star counts are versus targeted star counts\\     
(29)&W$\_$A,T$\_$b2$\_$U     &	The weight of blue box with $J2$ for UVES, where allocated star counts are versus targeted star counts\\     
(30)&W$\_$A,T$\_$b3$\_$U     &	The weight of blue box with $J3$ for UVES, where allocated star counts are versus targeted star counts\\     
(31)&W$\_$A,T$\_$b4$\_$U     &	The weight of blue box with $J4$ for UVES, where allocated star counts are versus targeted star counts\\       
(32)&W$\_$A,T$\_$e$\_$U	&	The weight of extra box for UVES, where allocated star counts are versus targeted star counts\\
(33)& Blue(\%)		     		&  The approximate fraction of targets in the blue versus the red box ($\%$)\\
\hline
\end{tabular}
\label{table:online_log}
\end{table*}

\section{Additional fields}\label{sec:AditionaFields}

A subset of fields were allocated to specially selected candidate targets belonging to rare but astrophysically important stellar populations, such as metal-poor stars or K~giants. Those additional fields in {\it{Gaia}}-ESO survey are labeled as Milky Way fields. The target selection for those additional fields is different from the Milky Way fields presented in this paper.
The additional fields were created with the target selection based on SkyMapper photometry \citep{keller07}, SDSS photometry \citep{ahn12}, VISTA VHS photometry and 2MASS photometry (see location of the fields on the sky in Fig.~\ref{fig:Mollweide}a).
In the following sections we present those different selections.

\subsection{Milky Way fields selected to study the other Galactic disc}\label{sec:sdss}

\begin{table}
\setlength{\tabcolsep}{.5pt}
\centering
\caption{Milky Way fields$^a$ for which SDSS photometry was used to select the targets to be observe with GIRAFFE (see Appendix~\ref{sec:sdss}).}
\begin{tabular}{lc}
\hline
Milky Way fields  &  $\Delta_{SDSS}$\\
\hline
\hline
GES$\_$MW$\_$082312-052959 & ... \\
GES$\_$MW$\_$083959-003000 & 0.41 \\
GES$\_$MW$\_$095600-003000 & ... \\
\hline
{\it{Note.}}$^a$For UVES targets the selection  \\ 
is based on 2MASS photometry.  
\end{tabular}
\label{table:sdss}
\end{table}

Three additional Milky Way fields were selected using SDSS photometry in order to study the outer disc of the Galaxy.
GIRAFFE fibres were allocated to candidate K giants, which probe the far outer disc, warp and flare.

The SDSS catalog flags adopted to select these Milky Way field targets are listed in Table \ref{tab:flags}.
The target selection of the additional fields in the blue and extra boxes with SDSS photometry is as follows:
\begin{equation}
\begin{split}
&{\rm Blue~box:} \\
& -0.3 \le (g-r) \le 1.0\\
&15.0 \le r \le 19.0 \\
&{\rm Extra~box:} \\
& -0.3 \le (g-r) \le 1.0+ \Delta_{SDSS} \\
&15.0 \le r \le 19.0 
\end{split}
\end{equation}
Here $\Delta_{SDSS}$ is the right-edge extension of the blue box. 
$\Delta_{SDSS}$ and the Milky Way field names selected with the SDSS photometry are listed in Table \ref{table:sdss}.
The selection of UVES targets in the same Milky Way fields is based on 2MASS photometry as described in Section~\ref{sec:UVES}.

\subsection{Milky Way fields with metal-poor stars in the halo}\label{sec:skymapper}

\subsubsection{Milky Way fields with the target selection based on SkyMapper photometry and 2MASS photometry}\label{sec:skymapper1}

Seven additional Milky Way fields were selected using SkyMapper photometry in order to study the metal-poor stars in the halo.

All GIRAFFE targets were selected from the SkyMapper photometry.
For the same Milky Way fields only one UVES fibre per field was devoted to observe a metal-poor star selected using SkyMapper photometry. 
The remaining UVES targets were selected using 2MASS photometry. 

Milky Way field names for which SkyMapper photometry was used to select the targets to be observe with GIRAFFE and one target to be observed with UVES are listed in Table~\ref{table:skymapper1}.

\subsubsection{Milky Way fields with the target selection based on SkyMapper photometry, 2MASS photometry and VISTA VHS photometry}\label{sec:skymapper2}
 
Table~\ref{table:skymapper2} list additional Milky Way fields selected to study the metal-poor stars in the halo.
One UVES target selected using SkyMapper photometry was dedicated to study the most interesting metal-poor star in the halo.
The remaining UVES targets were selected using 2MASS photometry.

For the GIRAFFE targets in the same Milky Way fields the selection is based on VISTA VHS photometry as described in Section~\ref{sec:Giraffe}.

\begin{table}
\setlength{\tabcolsep}{3.5pt}
\centering
\caption{Milky Way fields for which SkyMapper photometry was used to select the targets to be observe with GIRAFFE and one target to be observed with UVES$^a$ (see Appendix~\ref{sec:skymapper1}).}
\begin{tabular}{c}
\hline
\hline
Milky Way fields \\
\hline
\hline
GES$\_$MW$\_$094753-102657 \\
GES$\_$MW$\_$100913-412801 \\
GES$\_$MW$\_$101428-405235 \\
GES$\_$MW$\_$105731-124726 \\
GES$\_$MW$\_$105808-154324 \\
GES$\_$MW$\_$110053-132816 \\
GES$\_$MW$\_$131359-460007 \\
\hline
{\it{Note.}}$^a$The remaining UVES targets are selected \\ 
based on 2MASS photometry.  
\end{tabular}
\label{table:skymapper1}
\end{table}

\begin{table}
\setlength{\tabcolsep}{3.5pt}
\centering
\caption{Milky Way fields with the target selection based on SkyMapper$^a$ photometry, 2MASS$^b$ photometry and VISTA VHS$^c$ photometry (see Appendix~\ref{sec:skymapper2}).}
\begin{tabular}{c}
\hline
\hline
Milky Way fields\\
\hline
\hline
GES$\_$MW$\_$125609-451238\\
GES$\_$MW$\_$133026-434759\\
GES$\_$MW$\_$142145-440827\\
GES$\_$MW$\_$144113-400831\\
GES$\_$MW$\_$212402-431239\\
GES$\_$MW$\_$212731-542154\\
GES$\_$MW$\_$221259-455029\\
GES$\_$MW$\_$221818-582824\\
GES$\_$MW$\_$225008-554935\\
GES$\_$MW$\_$225108-524744\\
GES$\_$MW$\_$234854-560538\\
\hline
{\it{Note.}}
$^a$Only one SkyMapper target per field\\
was observed by UVES.\\
$^b$The remaining UVES targets were selected \\ 
based on 2MASS photometry.\\
$^c$For GIRAFFE targets the selection  \\ 
is based on VISTA VHS photometry. \\  
\end{tabular}
\label{table:skymapper2}
\end{table}


\bsp	
\label{lastpage}
\end{document}